\documentclass[a4paper,12pt]{article}
\usepackage[left=2cm,right=2cm,top=2cm,bottom=2cm]{geometry}
\usepackage[backend=bibtex,bibencoding=ascii,style=numeric,citestyle=numeric ,sorting=none]{biblatex}
\usepackage[utf8]{inputenc}
\usepackage[english]{babel}
\usepackage{csquotes}
\usepackage[T1]{fontenc}
\usepackage{amsmath}
\usepackage{amsfonts}
\usepackage{amsthm}
\usepackage{amssymb}
\usepackage{graphicx}
\usepackage{enumerate}
\usepackage{subfigure}
\usepackage{xcolor}
\usepackage{pbox}
\usepackage{lscape}
\usepackage{epstopdf}
\newtheorem{theorem}{Theorem}

\begin{document}
\title{An ecological resilience perspective on cancer: insights from a toy model}
\author{Artur C. Fassoni{$^{1}$} and Hyun M. Yang{$^{2}$}}
\date{}
\maketitle

\vspace{-1cm}
\begin{center}
\begin{small}
$^1$ fassoni[at]unifei.edu.br, IMC, UNIFEI, Itajubá, MG, Brazil \\
$^2$ hyunyang[at]ime.unicamp.br, IMECC, UNICAMP, Campinas, SP, Brazil
\end{small}
\end{center}

\bigskip

\begin{center}
\textbf{Abstract}
\end{center}

In this paper we propose an ecological resilience point of view on cancer. This view is based on the analysis of a simple ODE model for the interactions between cancer and normal cells. The model presents two regimes for tumor growth. In the first, cancer arises due to three reasons: a partial corruption of the functions that avoid the growth of mutated cells, an aggressive phenotype of tumor cells and exposure to external carcinogenic factors. In this case, treatments may be effective if they drive the system to the basin of attraction of the cancer cure state. In the second regime, cancer arises because the repair system is intrinsically corrupted. In this case, the complete cure is not possible since the cancer cure state is no more stable, but tumor recurrence may be delayed if treatment is prolongued. We review three indicators of the resilience of a stable equilibrium, related with size and shape of its basin of attraction: latitude, precariousness and resistance. A novel method to calculate these indicators is proposed. This method is simpler and more efficient than those currently used, and may be easily applied to other population dynamics models. We apply this method to the model and investigate how these indicators behave with parameters changes. Finally, we present some simulations to illustrate how the resilience analysis can be applied to validated models in order to obtain indicators for personalized cancer treatments.

\bigskip


\section{Introduction}

The ecological resilience perspective is an emerging approach for understanding the dynamics of social-ecological systems \cite{holling1973resilience,may1977thresholds,scheffer2001catastrophic, folke2006resilience, menck2013basin, meyer2016mathematical}. While the stability point of view emphasizes the equilibrium and the maintenance of present state, the resilience point of view focus on shifts between alternative basins of attraction, thresholds, uncertainty and unexpected disturbances. External forces or random events may cause state variable perturbations that drive a nonlinear system, which is initially near a stable state, to enter an undesirable basin of attraction. In this case, the resilience of the original steady state is related with the size and shape of its basin of attraction, and the capacity of the system to persist in this basin of attraction when subject to state variable perturbations. Three different indicators are established in the literature as measures of the resilience of a stable state with respect to state variable perturbations \cite{walker2004resilience,mitra2015integrative}: \textit{latitude}, which is a measure of the volume of its basin of attraction; \textit{precariousness}, which is related with the minimal state space disturbance needed to drive the system outside its basin of attraction; and \textit{resistance}, which is a measure of the deepness of its basin of attraction.

On the other hand, changes in system parameters occur in a slow time scale, due to evolutionary forces or by modifying the intensities of interactions and forces governing such system. In this case, parameters modify the resilience of the system with respect to state variable perturbations. Further, when parameters do change enough, the system may undergo several bifurcations and the phase portrait may change substantially. In this case, one can measure the resilience of the system with respect to parameters changes as the distance to the threshold values for which bifurcations occur. As a consequence of such bifurcations, an undesirable alternative stable state may be created, and its basin of attraction can be achieved by state variable perturbations, as commented above. A more dramatic outcome happens when parameters changes lead to loss of stability of the original steady state or even its disappearance. In this case, a regime shift occurs and the system moves to another state. Now, the question of reversibility takes place. Of first importance is the question whether it is possible or not to return the parameters to their original values. When parameters change due to evolutionary factors, it is more likely that this change can not be undone. Changes due to external forces can be undone more easily through the correct manipulation of those forces (if possible). However, even if the original values can by restored, the reversal to the original stable state may not be completely achieved if the system exhibits hysteresis.

In this paper we illustrate how these concepts of ecological resilience can be applied to cancer, a complex disease whose causes are far from being well understood and whose cure is far from being achieved. Indeed, despite the intense efforts that led the elucidation of many biochemical mechanisms developed by cancer cells to survive \cite{hanahan2011hallmarks}, there is a current debate on which are the major factors that allow the onset of cancer cells. While some argue that alterations in intrinsic cellular processes are the main reasons that some tissues become cancerous \cite{tomasetti2015variation}, others defend the view that most cases of cancer result from extrinsic factors such as environmental exposure to toxic chemicals and radiation \cite{wu2016substantial}. With respect to cancer treatment, although the development of new drugs and strategies to treat cancer in the last fifty years achieved good results in many cases, another large portion of cancer patients did not respond well to treatments, or presented tumor recurrence, indicating that there is still a long road in the fight against cancer \cite{kerbel2004anti,benzekry2015metronomic}.

We propose a simple model for tumor growth and apply the above concepts to suggest a framework for viewing the arising of cancer and its effective treatment as critical transitions between two alternative stable states. In this framework, tumor growth and tumor treatment depend ultimately on ecological resilience questions. Further, we briefly review the three resilience indicators commented above, propose a method to calculate these indicators and apply this method to the model. As far as we know, this novel method we propose is simpler and more efficient than those currently used, and can be applied to other population dynamics models to improve their analysis through this resilience perspective. 

The paper is organized as follows. In Section \ref{secNA:model} the model is presented. In Section \ref{sec:Anal} the analysis of the model is performed. In Section \ref{secNA:ecol}, the results are discussed in the ecological resilience perspective. In Section \ref{secNA:res}, the method to calculate resilience indicators is presented and applied to the model. Finally, conclusions are presented in Section \ref{secNA:con}.

\section{A toy model for tumor growth}
\label{secNA:model}

We present a simple model consisting of a system of ODEs describing tumor growth and its effect on normal tissue, together with the tissue response to tumor. Our goal is not to consider the several aspects of tumor growth and to reproduce quantitative behavior with high accuracy, but to use the model to give some insights through a resilience point of view. The model equations are given by
\begin{subequations}
\label{sisNA}
\begin{align}
 \dfrac{dN}{dt}= &\; r_N - \mu_N N - \beta_1 N A,
\label{dNdt_NA} \\
 \dfrac{dA}{dt}= &\; r_A A \left(1-\frac{A}{K_A}\right)  - \beta_3 N A - (\mu_A+\epsilon_A) A,
\label{dAdt_NA}
\end{align}
\end{subequations}
where $N$ and $A$ stand for normal and tumor cells, respectively. This system is a limit case of a three-dimensional model for oncogenesis encompassing mutations and genetic instability \cite{fassoni2016thesis}.

Parameter $r_N$ represents the total constant reproduction of normal cells, and $\mu_N$ is their natural mortality. A constant flux for normal cells is considered in the vital dynamics, and not a density-dependent one, like the logistic growth generally assumed \cite{gatenby1995models, gatenby1996reaction, de2001mathematical, mcgillen2014general}. The reason for this choice is that at a normal and already formed tissue the imperative dynamics is not the cells intraspecific competition by nutrients, but the maintenance of a homeostatic state, through the natural replenishment of old and dead cells \cite{simons2011strategies}.

On the contrary, cancer cells have a certain independence on growth signals released by the tissue and keep their own growth program, like an embrionary tissue in growth phase \cite{fedi1997growth}. Thus a density dependent growth is considered. Several growth laws could be used, such as the Gompertz, generalized logistic, Von Bertanlanfy and others \cite{sarapata2014comparison}. We choose the logistic growth due to its simplicity, and a natural mortality $\mu_A$. An extra mortality rate $\epsilon_A$ due to apoptosis \cite{danial2004cell} is also included.

Several models for tumor growth consider the phenomena of tumor angiogenesis, i.e., the formation of new blood vessels to feed the tumor, in response to signals released by tumor cells \cite{kerbel2008tumor,yang2012mathematical}. In order to keep the model as simple as possible, we do not consider angiogenesis here. 

Parameter $\beta_3$ encompasses, in the simplest way, all negative effects imposed to cancer cells by the many cell types in the normal tissue. These interactions include the release of anti-growth and death signals by host cells \cite{hanahan2011hallmarks}, the natural response of normal cells to the presence of cancer cells, the competition by nutrients with tumor cells and so on. Similarly, parameter $\beta_1$ covers all mechanisms developed by tumor cells which damage the normal tissue, like increasing local acidity \cite{gatenby2006acid}, supression of immune cells \cite{facciabene2012t}, release of death signals \cite{hanahan2011hallmarks}, and competition with normal cells.

System \eqref{sisNA} is similar to the classical Lotka-Volterra model of competition \cite{fassoni2014basins}, commonly used in models for tumor growth \cite{gatenby1995models, gatenby1996reaction, de2001mathematical, mcgillen2014general} and biological invasions \cite{fassoni2014mathematical}, but has a fundamental difference. The use of a constant flux instead a logistic growth to normal cells breaks the symmetry observed in the classical Lotka-Volterra model, so that no equilibrium with $N=0$ will exist. Thus, normal cells will never be extinct, on the contrary to those models. We believe that this is not a problem, but, on the contrary, is a realistic outcome. Indeed, roughly speaking, cancer `wins' not by the fact that it kills all cells in the tissue, but by the fact that it reaches a dangerous size that disrupts the well functioning of the tissue and threaten the health of the individual. A constant flux term was already taken in other well-known models for cancer, specifically, to describe the growth of immune cells \cite{kuznetsov1994nonlinear, de2005validated, eftimie2011interactions}.

Let us comment on some resemblance of system \eqref{sisNA} with the well-known system of Kuznetsov et. al \cite{kuznetsov1994nonlinear}, which describes the interaction between immune cells and cancer cells. In that system, equation for immune cells has two production terms: a constant production term (analog to $r_N$ here) and a Michaelis-Menten term representing the recruitment of immune cells due to the presence of cancer cells. If we remove this second term (letting $p=0$ in their notation), that system becomes equivalent to system \eqref{sisNA}. Thus, the immune cells of that model have basically the same behavior of normal cells in this model, and the unique difference is the recruitment term. In our model, as the population $N$ is considered as a pool of many different cell types, from which the immune cells are a small fraction, it is natural to include in its dynamics only the common behavior of all these types, disregarding the particularities of each type.

\section{Analysis of the model}
\label{sec:Anal}

We now present the analysis of system \eqref{sisNA}. Biological implications are discussed in Section \ref{secNA:ecol}.

\subsection{Equilibrium points}
System \eqref{sisNA} has a trivial equilibrium
\begin{equation*}
P_0=\left(\dfrac{r_N}{\mu_N},0\right),
\end{equation*}
and up to two nontrivial equilibria
\[
P_i=({N}_i,{A}_i)=\left(\frac{r_N}{\mu_N+\beta_1 {A}_i} , {A}_i\right), \ i=1,2.
\]
Here, ${A}_1$ and $A_2$ are the roots of the second degree equation
\begin{equation}
aA^2+b A+c=0,
\label{qA_NA}
\end{equation}
with coefficients
\begin{equation*}
a= \dfrac{\beta_1 r_A}{K_A}>0,
\,\,\,\,\,\,\,\,\,\,\,\,
b=l_A\left(\beta_1^{th} - \beta_1\right),
\,\,\,\,\,\,\,\,\,\,\,\,
c=r_N \left(\beta_3 - \beta_3^{th}\right).
\end{equation*}
where
\begin{equation}
\beta_1^{th}=\frac{\mu_N r_A}{l_A K_A},
\label{b1b3th}
\ \  \ \
\beta_3^{th} =\frac{\mu_N}{r_N}l_A,
\ \ \ \ {\rm and} \ \ \ \ l_A=r_A-(\mu_A+\epsilon_A).
\end{equation}
When $A_1$ and $A_2$ are real, we label them in the order ${A}_1<{A}_2$. Conditions for having positive equilibria ${P}_1$ and ${P}_2$ are obtained by Descartes' Rule of Signs. Together with the trivial equilibrium $P_0$, the results are summarized as follows:
\begin{enumerate}
\item[I)] If $\beta_3>\beta_3^{th}$ and $\beta_1<\beta_{1,\Delta}^{th}$, the unique nonnegative equilibrium is the trivial equilibrium $P_0$.
\item[II)] If $\beta_3>\beta_3^{th}$ and $\beta_1>\beta_{1,\Delta}^{th}$, three nonnegative equilibria are $P_0$, ${P}_1$ and ${P}_2$.
\item[III)] If $\beta_3<\beta_3^{th}$, the nonnegative equilibria are $P_0$ and ${P}_2$.
\end{enumerate}
The threshold $\beta_{1,\Delta}^{th}$, defined for $\beta_3>\beta_3^{th}$, is the value of $\beta_1>\beta_1^{th}$ for which the discriminant $\Delta=b^2-4ac$ is zero, and is given by
\begin{equation}
\beta_{1,\Delta}^{th}= \beta_1^{th} + 2\eta + 2 \sqrt{\eta(\beta_1^{th}+\eta)}, 
\label{b1thD}
\end{equation}
where $\eta = r_A r_N (\beta_3 - \beta_3^{th})/(K_Al_A^2)$.

The threshold $\beta_3^{th}$ can be written as $\beta_3^{th} =l_A/N_0$, where $N_0=r_N/\mu_N$ is the number of normal cells in the absence of tumor cells. As we will see, condition $\beta_3>\beta_3^{th}$ implies on more difficult regimes for cancer to arise. Thus, favorable regimes for tumor growth occur for larger values of net reproduction rate $l_A$ of tumor cells and for smaller values of carrying capacity of normal cells $N_0$. The threshold $\beta_1^{th}$ for tumor aggressiveness increases as the mortality $\mu_N$ of normal cells increases, and decreases as the effective carrying capacity of tumor cells $l_AK_A/r_A$ increases.

\subsection{Local Stability}
\label{stabNA}

Stability of $P_0$ is easily determined. The eigenvalues of the Jacobian matrix of system \eqref{sisNA} evaluated at $P_0$ are given by
\[
\lambda_{1}=-\mu_N, \,\,\, \text{and} \,\,\,  \lambda_{2}=\frac{r_N}{\mu_N}(\beta_3^{th}-\beta_3).
\]
Thus, $P_0$ is locally asymptotically stable if $\beta_3>\beta_3^{th}$, and is a saddle otherwise.

We now study the local stability of ${P}_i$, $i=1,2$. Using the fact that ${N}_i$ and ${A}_i$ satisfy
\begin{equation}
\label{eqNApbarra}
l_A - \frac{r_A}{K_A} {A}_i - \beta_3 {N}_i =0,
\end{equation}
we find that the Jacobian matrix of system \eqref{sisNA} evaluated at ${P}_i$ is given by
\begin{equation*}
J({P}_i) = \left[ \begin{array}{cc}
-\beta_1 {A}_i - \mu_N & -\frac{\beta_1}{\beta_3}(l_A - \frac{r_A}{K_A} {A}_i) \\
-\beta_3 {A}_i & -\frac{r_A}{K_A} {A}_i
\end{array} \right].
\label{jacobNA}
\end{equation*}
Whenever ${P}_i$ is a positive equilibrium, the trace of $J({P}_i)$ is negative. Thus, when ${P}_i$ is positive, both eigenvalues of $J({P}_i)$ will have negative real part if $\mathrm{det}(J({P}_i))>0$, and we have opposite signs in the other case. Calculating the determinant we obtain
\begin{equation*}
\mathrm{det}(J({P}_i))= {A}_i\left(2\dfrac{\beta_1 r_A}{K_A}{A}_i+\dfrac{\mu_Nr_A}{K_A}-\beta_1l_A\right)= {A}_i(2a{A}_i+b),
\label{detjpbarra}
\end{equation*}
where $a$ and $b$ are the coefficients of \eqref{qA_NA}. We have that ${A}_{1,2}=(-b\pm\sqrt{\Delta})/2a$, with ${A}_1<{A}_2$, where $\Delta$ is the discriminant of \eqref{qA_NA}. Thus, whenever ${P}_i$ is a positive equilibrium, $i=1,2$,
\begin{equation*}
\mathrm{det}(J({P}_1))=-{A}_1\sqrt{\Delta}<0,
\,\,\,\,\,\,\,\,
\text{ and }
\,\,\,\,\,\,\,\,
\mathrm{det}(J({P}_2))={A}_2\sqrt{\Delta}>0.
\end{equation*}
Thus, ${P}_1$ will be a saddle point whenever it is positive (case II above), and ${P}_2$ will be stable whenever it is positive (cases II and III above).

\subsection{Asymptotic behavior and global stability}

Let us show the boundedness of trajectories of \eqref{sisNA}. By noting that
\begin{equation*}
\dfrac{dN}{dt} \leq r_N - \mu_N N,
\end{equation*}
and
\begin{equation*}
\dfrac{dA}{dt}\leq l_A A \left(1-\frac{r_A}{l_AK_A}A\right),
\end{equation*}
we may apply classical comparison principles \cite{cantrell1996practical} and conclude that all solutions $(N(t),A(t))$ with non-negative initial values remain restricted in the box 
\begin{equation}
B=\left[0,\frac{r_N}{\mu_N}\right] \times \left[0,\frac{l_A }{r_A}K_A\right]
\label{boxB}
\end{equation}
when $t\to \infty$. 

In order to rule out periodic orbits for system \eqref{sisNA} we apply the Dulac Criterion \cite{strogatz2001nonlinear} with $u(N,A)=1/NA$, obtaining
\[
\nabla \cdot \left(\frac{1}{NA}\left(\dfrac{dN}{dt},\dfrac{dA}{dt}\right) \right) = - \frac{r_A NA+K_Ar_N}{K_A A N^2} <0 
\]
for $(N,A)\in B$. Thus, system \eqref{sisNA} has no periodic orbits.

By the Poincar\'e-Bendixson Theorem we conclude that all trajectories converge to an equilibrium point \cite{strogatz2001nonlinear}. It implies that equilibria $P_0$ and ${P}_2$ are globally stable in cases I and III, respectively (in the latter, $P_2$ is globally stable for initial conditions $A(0)>0$, since the $N$ axis is the stable manifold of $P_0$). In case II, the plane $N\times A$ is divided in the basins of attraction of $P_0$ and ${P}_2$. The stable manifold of ${P}_1$ is the separatrix between these basins. All these results are summarized in Theorem \ref{thm_NA_analise}.

\begin{theorem}
\label{thm_NA_analise}
System \eqref{sisNA} has the following behavior:
\begin{enumerate}
\item[I)] If $\beta_3>\beta_3^{th}$ and $\beta_1<\beta_{1,\Delta}^{th}$, then $P_0$ is globally stable.
\item[II)] If $\beta_3>\beta_3^{th}$ and $\beta_1>\beta_{1,\Delta}^{th}$, then $P_0 $ and ${P}_2$ are locally stable. Equilibrium ${P}_1$ is a saddle point whose stable manifold is the separatrix between the basins of attraction of $P_0$ and ${P}_2$.
\item[III)] If $ \beta_3<\beta_3^{th} $, then ${P}_2$ is globally stable for initial conditions $A(0)>0$.
\end{enumerate}
\end{theorem}

\begin{center}
\begin{figure}[!htb]
\begin{minipage}[b]{\linewidth}
\begin{center}
\includegraphics[width=0.45\linewidth]{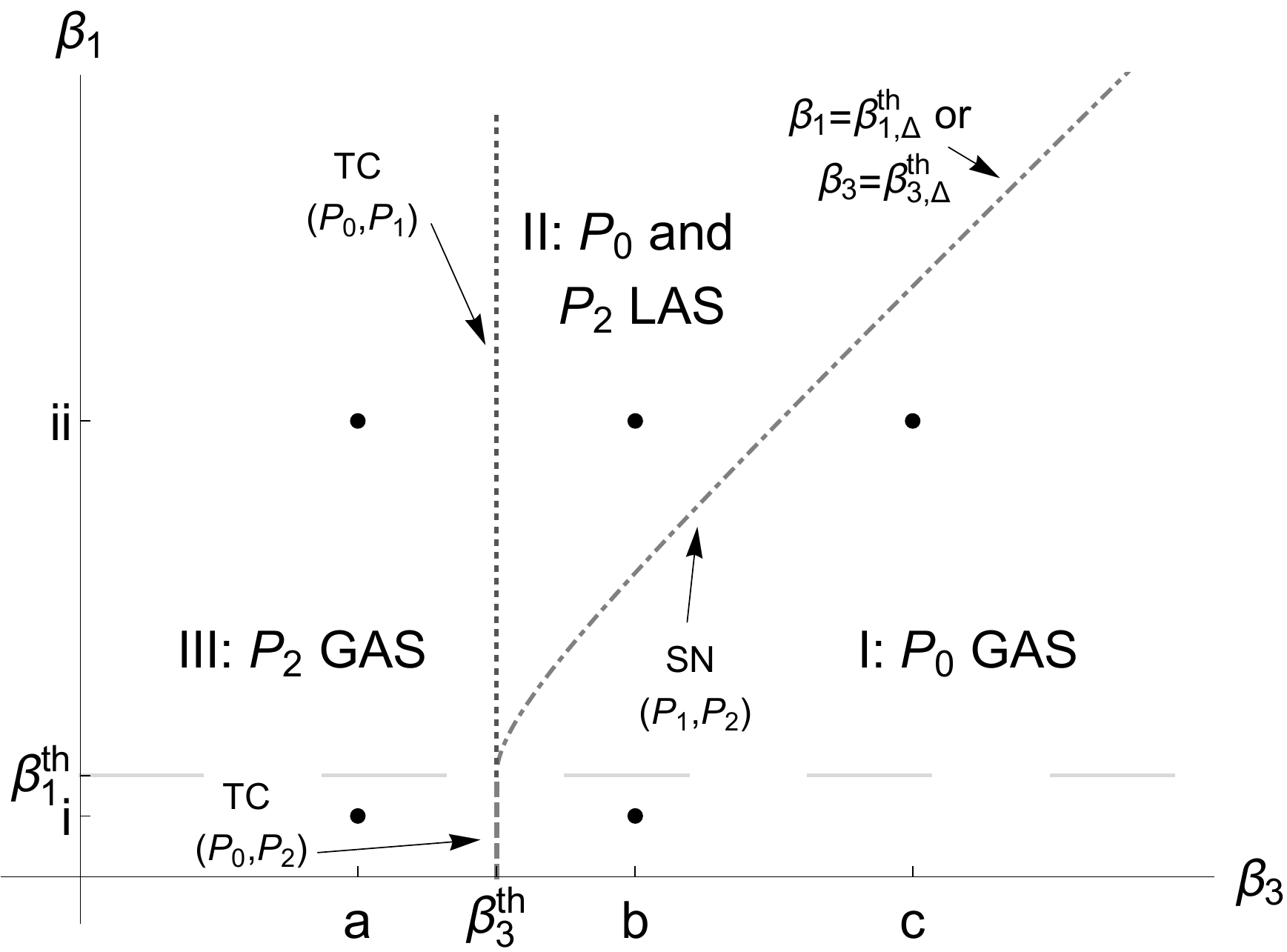}
\hfill
\includegraphics[width=0.45\linewidth]{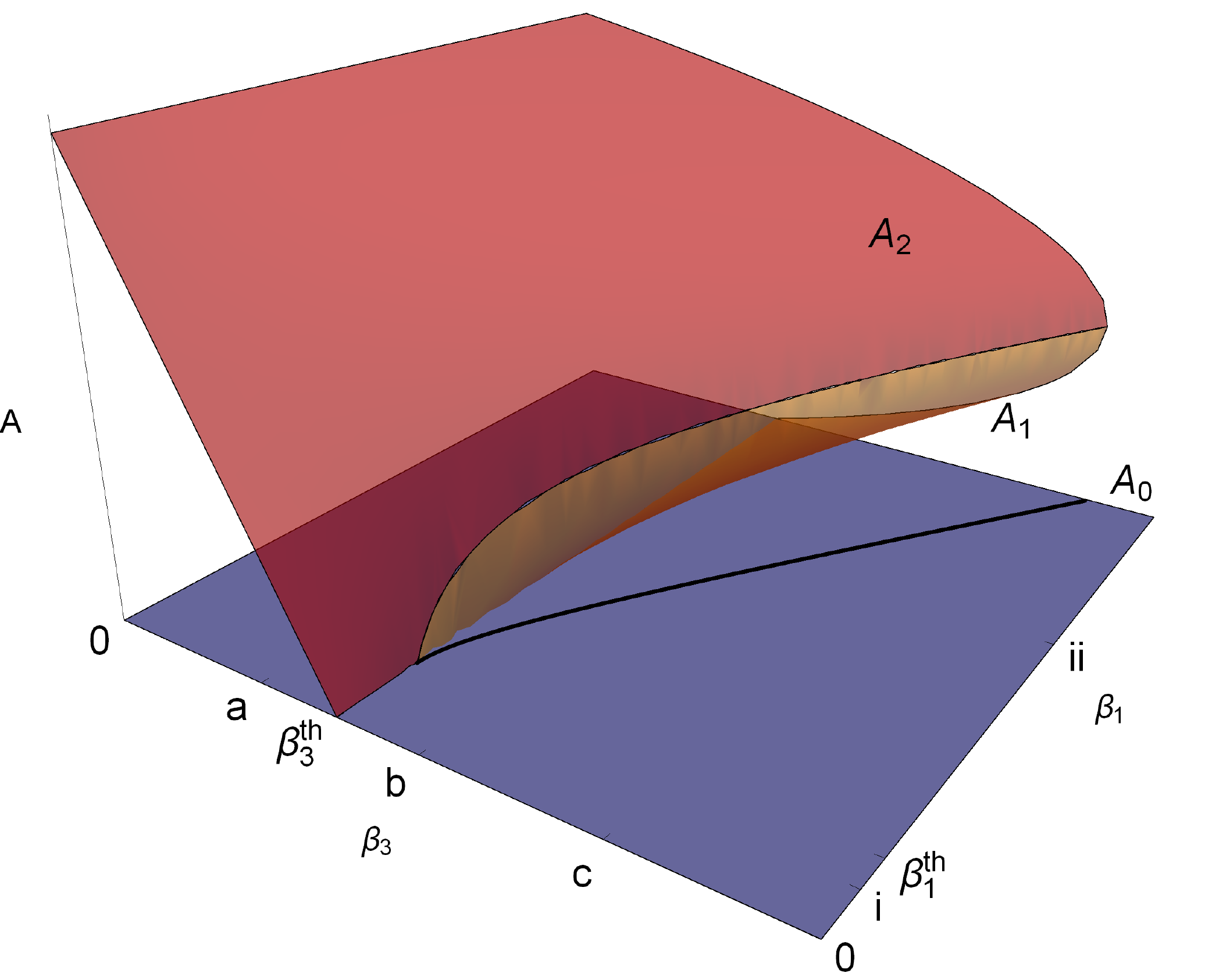}
\end{center}
\end{minipage}
\caption{
Left: Regions I, II and III in the $\beta_3 \times \beta_1$ plane of parameters space, together with equilibria behavior. If $\beta_1<\beta_{1}^{th}$, a direct transition from III to I occurs as $\beta_3$ increases, with a transcritical bifurcation between $P_0$ and $P_2$ when $\beta_3=\beta_3^{th}$. If $\beta_1>\beta_{1}^{th}$, a transition from III to II and a transcritical bifurcation between $P_0$ and $P_1$ occurs at $\beta_3=\beta_3^{th}$; a second transition, from II to I, occurs when $\beta_3=\beta_{3,\Delta}^{th}$, with a saddle-node bifurcation between $P_1$ and $P_2$ (for an expression and details about $\beta_{3,\Delta}^{th}$, see Section \ref{secNA:ecol}.1). Right: bidimensional diagram of the $A$ coordinates of equilibria $P_2$ (red, stable), $P_1$ (orange, unstable) and $P_0$ (blue, stable for $\beta_3>\beta_3^{th}$), when $\beta_3 $ and $ \beta_1$ vary. 
}
\label{fig-bif-b1b3}
\end{figure}
\end{center}

The division of the $\beta_1 \times \beta_3$ plane into regions I, II and III is showed in Figure \ref{fig-bif-b1b3}.

\subsection{Numerical simulations}

\begin{center}
\begin{figure}[htb!]
\begin{minipage}[b]{\linewidth}
\begin{center}
\includegraphics[width=.47\linewidth]{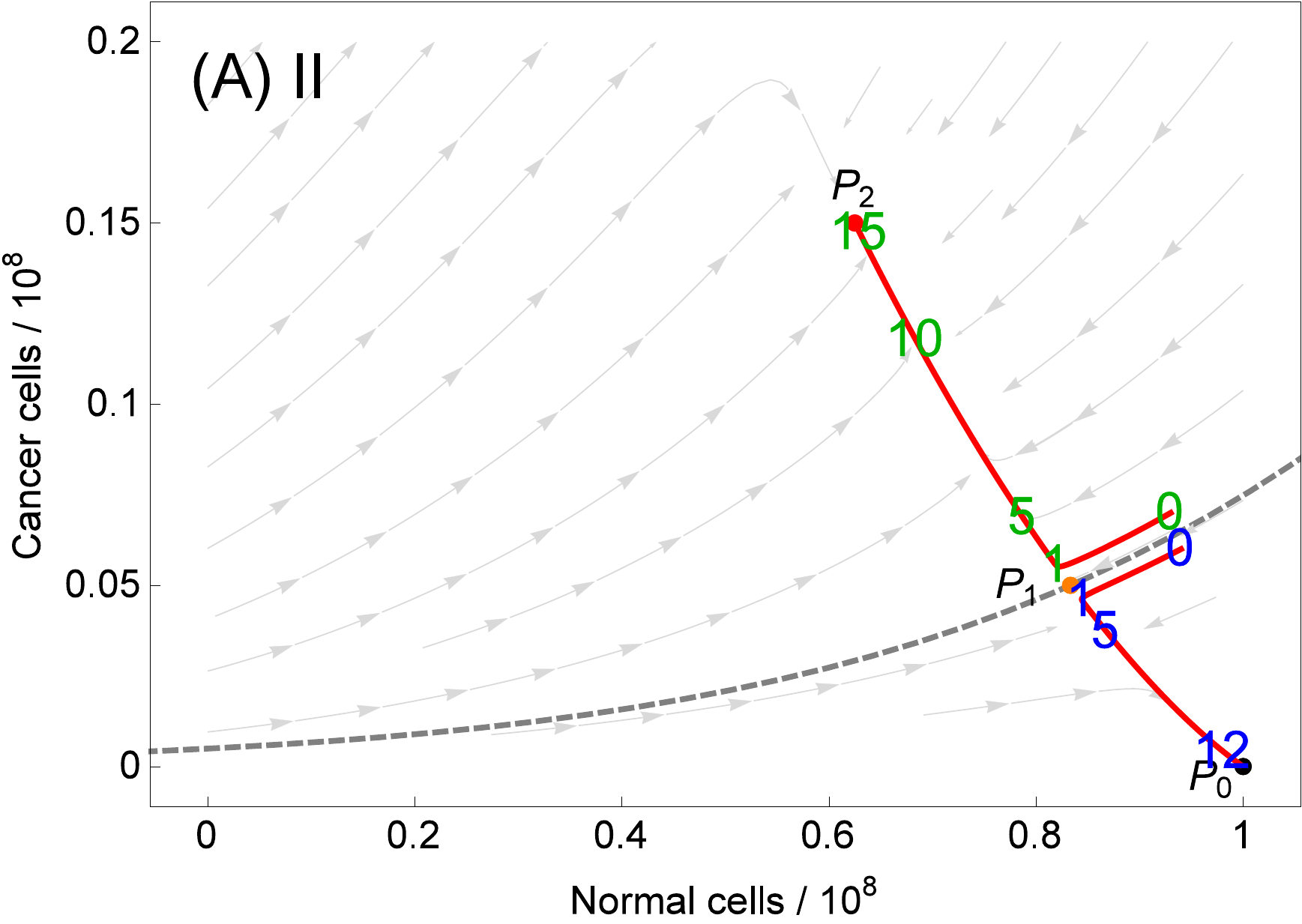}
\hfill
\includegraphics[width=.47\linewidth]{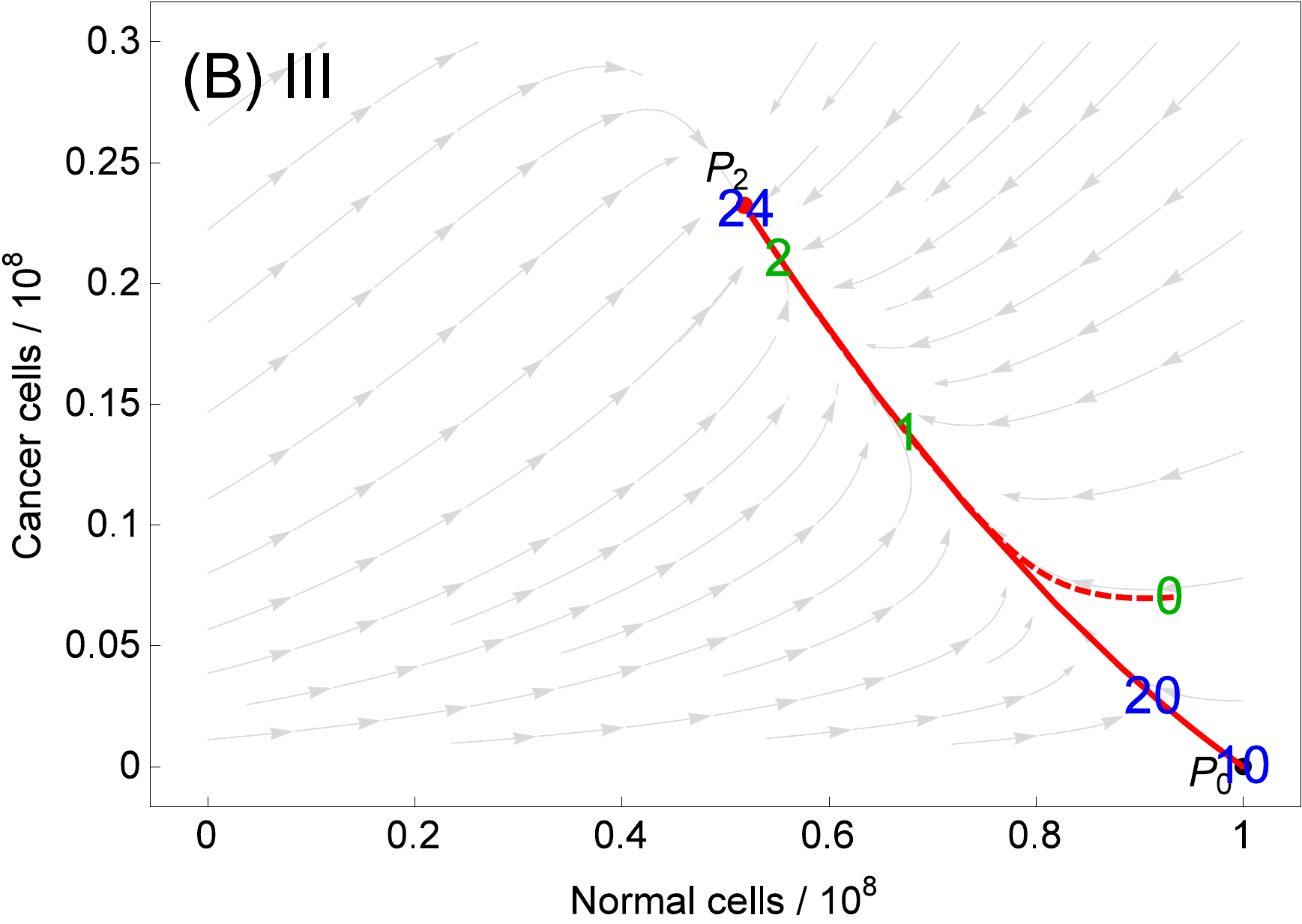}
\end{center}
\end{minipage}
\caption{Solutions os system \eqref{sisNA} when parameters correspond to regimes II (A) and III (B), using values in Table \ref{tablePar}. Initial conditions are $N(0)=r_N/\mu_N-A_0$ and $A(0)=A_0$, representing that the tissue was initially at its homeostatic state when $A_0$ cells have become cancerous. The values of $A_0$ are: (A) $A_0=0.06\times 10^8$ cells and $A_0=0.07\times 10^8$ cells; (B) $A_0=1$ cell and $A_0=0.07\times 10^8$ cells. In the left panel, the gray dotted curve represents the separatrix between the basins of attraction of $P_0$ and $P_2$. The green and blue numbers indicate the time (in years) corresponding to the trajectory.}
\label{fig_NA_onset}
\end{figure}
\end{center}

We now present numerical simulations in order to stimulate discussions in next sections. Figure \ref{fig_NA_onset} shows simulations of system \eqref{sisNA} in cases II and III. Parameters values were based on data from the literature, specially for breast cancer, according to the procedure below. A summary of the parameter values is presented in Table \ref{tablePar}.

\begin{table}[htb!]
\begin{center}
\caption{Parameters description and values adopted in simulations.\label{tablePar}}
\begin{tabular}{cllll}
\hline 
Parameter & Description &  Value 
\\  \hline 
$\mu_N$ & $1/\mu_N$ is the lifetime of a normal cell 
& $0.01$ day$^{-1}$
\\ 
$r_N$ & total constant reproduction of normal cells
& $10^6$  cell day$^{-1}$ 
\\ 
$r_A$ &  tumor cells growth rate
& $0.05$ day$^{-1}$
\\ 
$K_A$ & tumor carrying capacity 
& $0.75 \times 10^8$ cells
\\ 
$\mu_A$ & natural mortality rate of cancer cells
& $0.01$ day$^{-1}$
\\ 
$\epsilon_A$ & extra mortality rate of cancer cells
& $0.01$ day$^{-1}$
\\ 
$\beta_1$ & cancer cells aggressiveness
&  $ 0.40 \times 10^{-9}$ cell$^{-1}$day$^{-1}$
\\ 
$\beta_3^{II}$ & tissue response to cancer cells - case II
&  $ 0.28 \times 10^{-9}$ cell$^{-1}$day$^{-1}$ 
\\ 
$\beta_3^{III}$ & tissue response to cancer cells - case III 
&  $ 0.32 \times 10^{-9}$ cell$^{-1}$day$^{-1}$ 
\\
\hline 

\end{tabular}
\end{center}
\end{table}

We assume that the lifetime of a normal cell is $100$ days, thus $\mu_N = {1}/{100}$ days$^{-1}$. The number of normal cells in the breast cannot pass $N_0=10^8$ cells \cite{spencer2004ordinary}. Thus, in order to adjust the equilibrium ${r_N}/{\mu_N}$ of normal cells in the absence of cancer to be $10^8$ cells, we consider $r_N=10^6$ cells/day. For cancer cells, we assume the same natural mortality, $\mu_A={1}/{100}$ days$^{-1}$. For the apoptotic rate of $A$ cells, we use $\epsilon_A=1/100$ days$^{-1}$. The ratio \textit{birth rate}/\textit{death rate} is $1$ for a normal tissue, in order to maintain a homeostatic state. Following \cite{spencer2004ordinary}, we assume that cancer cells have increased this ratio by a factor of five. Thus, we have $r_A={5}/{100}$ days$^{-1}$. In order to have the maximum number of cancer cells being 75$\%$ of the normal cells, we consider $K_A=7.5 \times 10^7$. The values of interacting parameters $\beta_1$ and $\beta_3$, in units of cell$^{-1}$day$^{-1}$, are unknown a priori but, by substituting the values of other parameters, we obtain the thresholds for $\beta_1$ and $\beta_3$: the threshold $\beta_3^{th}$ which separates cases II and III has the value $\beta_3^{th} = 0.30 \times 10^{-9}$. So, we assume two possible values for $\beta_3$: $\beta_3^{III} = 0.28 \times 10^{-9}$, and $\beta_3^{II} = 0.32 \times 10^{-9}$. Each of these values will originate a different behavior of system \eqref{sisNA} (see Figure \ref{fig-bif-b1b3}). If $\beta_3=\beta_3^{III}$ we are in case III, for every value of $\beta_1$. If $\beta_3=\beta_3^{II}$, we have $\beta_{1,\Delta}^{th} = 0.37 \times 10^{-9}$, so we assume $\beta_1=0.40 \times 10^{-9}>\beta_3^{II}$, which is reasonable since cancer cells are supposed to cause more damage to normal cells than the contrary. With these values we are in case II. In all numerical simulations in this paper, we use these parameter values, and $\beta_3=\beta_3^{III}$ or $\beta_3=\beta_3^{II}$, depending on the interest to simulate cases III or II.

\section{An ecological resilience perspective on cancer}
\label{secNA:ecol}

We now discuss the biological implications of the previous analysis. Our look to system \eqref{sisNA} as being a simple cartoon, a toy model of the underlying system governing tumor growth in a cancer patient and thus we apply the perspective of ecological resilience to discuss the above results. Although this is a rough approximation, it may be instructive illustration on our understanding of cancer onset and cancer treatment. By cartoon or an approximation, we mean that the underlying system of cancer in real life, despite being very complex, may present three qualitative distinct regimes, corresponding to regimes I, II and III of system \eqref{sisNA}. In this analogy, an equilibrium state corresponding to the presence of a tumor is not necessarily a static equilibrium, but a state of the system where a tumor is growing and developing. Let us discuss the differences between these regimes. 

\subsection{Cancer onset as a critical transition}

Initially, we look to cancer onset as a critical transition. Let us first comment on the `natural repair system of the patient', a mechanism which is operated at a variety of levels and by many agents. In the tissue level, it is operated by the immune system, through lymphocytes and natural killer cells for example \cite{vivier2012targeting}. The presence of cancer cells at a given site stimulate the locomotion of immune system cells to that site in order to eliminate the cancer cells. At the cellular level, many cell components watch some parameters of the own cell and its neighbors, as DNA integrity, the products of cellular metabolism, the concentration of growth factors, etc. When abnormal conditions are detected, the cell may kill itself through apoptosis \cite{danial2004cell}, or the normal cells may release death or inhibitor factors to control the undesired growth \cite{hanahan2011hallmarks}. In system \eqref{sisNA}, these mechanisms are roughly described by parameters $\beta_3$ and $\epsilon_A$. These are the parameters most subject to changes in a slow-time scale, through the multistep process of genetic alterations which transform the descendants of a normal cell into a malign tumor \cite{hanahan2011hallmarks}. Parameter $\beta_1$ is also thought to be a varying parameter in this slow-time scale, since it encompasses the many types of negative interactions which cancer cells impose to the host tissue, especially due to changes in their metabolism which increase local acidity or lead to starvation of oxygen and nutrients for normal cells.

In the first regime (region I), we have a healthy person, since $P_0$ is globally stable. In this case, we have an efficient tissue repair system, since condition $\beta_3>\beta_3^{th}$ can be written as
\[
\beta_3 + \epsilon_A \frac{\mu_N}{r_N} > \frac{\mu_N}{r_N}(r_A-\mu_A).
\]
Further, we have a limited aggressiveness of cancer cells, $\beta_1<\beta_{1,\Delta}^{th}$. This condition depends also on $\beta_3$, because $\beta_{1,\Delta}^{th}$ depends on $\beta_3$ (see Figure \ref{fig-bif-b1b3}, left). Therefore, for a fixed $\beta_1$, condition $\beta_1<\beta_{1,\Delta}^{th}$ is equivalent to $\beta_3>\beta_{3,\Delta}^{th}$, where $\beta_{3,\Delta}^{th}$ is the inverse function of $\beta_{1,\Delta}^{th}$. Thus, for each level of aggressiveness of cancer cells, i.e., for each fixed $\beta_1$, we have a second threshold, $\beta_{3,\Delta}^{th}$, that the tissue response $\beta_3$ must be above in order to completely eliminate the chance of cancer. The more aggressive are the cancer cells, the higher is the threshold  $\beta_{3,\Delta}^{th}$. Thus, region I corresponds to parameters such that, although new mutant cells may arise all the time, they are not so much aggressive and the intrinsic repair system is capable to eliminate them.

In the second regime (region II), we have the possibility of having cancer, since $P_0$ and $P_2$ are both stable. Condition $\beta_3>\beta_3^{th}$ implies that the tissue response is efficient, but condition $\beta_1>\beta_{1,\Delta}^{th}$, which is equivalent to $\beta_3<\beta_{3,\Delta}^{th}$, implies that this response is not completely capable to face the aggressiveness of cancer cells. Thus, region II corresponds to a partially corrupted repair system due to the aggressiveness of cancer cells. In this region, the resilience of the cancer cure equilibrium $P_0$ plays an important role. The survival and installation of a tumor mass depend on factors which favor the mutations from normal to cancer cells, such as increased genetic instability and/or exposure to external carcinogenic agents \cite{negrini2010genomic}. These factors can lead to an increase in the initial conditions of cancer cells and allow them to surpass the threshold separating the basins of attraction of $P_0$ and $P_2$ (see Figure \ref{fig_NA_onset} left). Therefore, it is important to analyze how the resilience of $P_0$ behave as key parameters are changed. In the next Section we develop this `resilience analysis'. We refer to \cite{fassoni2016thesis} for a model that comprises an intermediary pre-cancer population that continuously `feed' the population $A$ with new individuals.

Finally, case III represents a dramatic corruption of the repair system, $\beta_3<\beta_3^{th}$. Now, the cancer equilibrium $P_2$ is globally stable, and the onset and development of cancer are possible for any initial number of cancer cells. However, distinct quantitative behaviors may be observed, depending on the initial number of cancer cells. In Figure \ref{fig_NA_onset} we see that the time at which the tumor reaches a detectable size (approximately $10^6$ cells \cite{schabel1975concepts}) is more than ten years if a single mutant cell arises. As the number of initial cancer cells increases, this time decreases.

\begin{center}
\begin{figure}[!htb]
\begin{minipage}[b]{\linewidth}
\begin{center}
\includegraphics[width=0.47\linewidth]{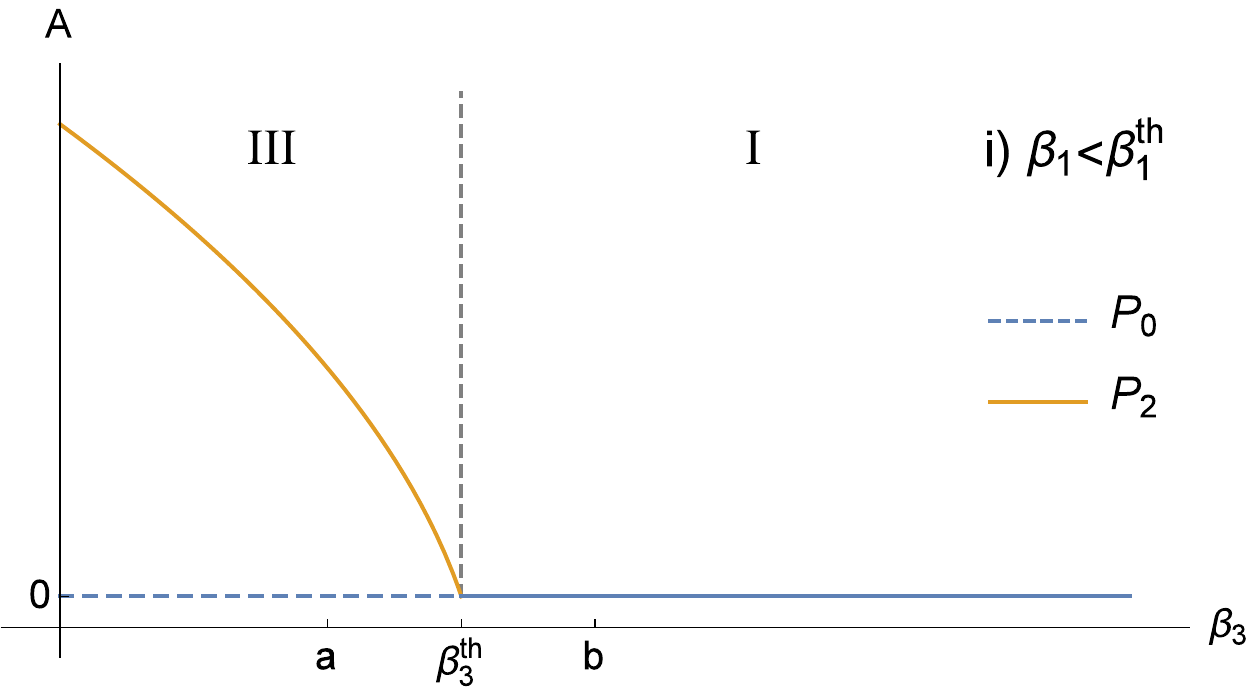}
\hfill
\includegraphics[width=0.47\linewidth]{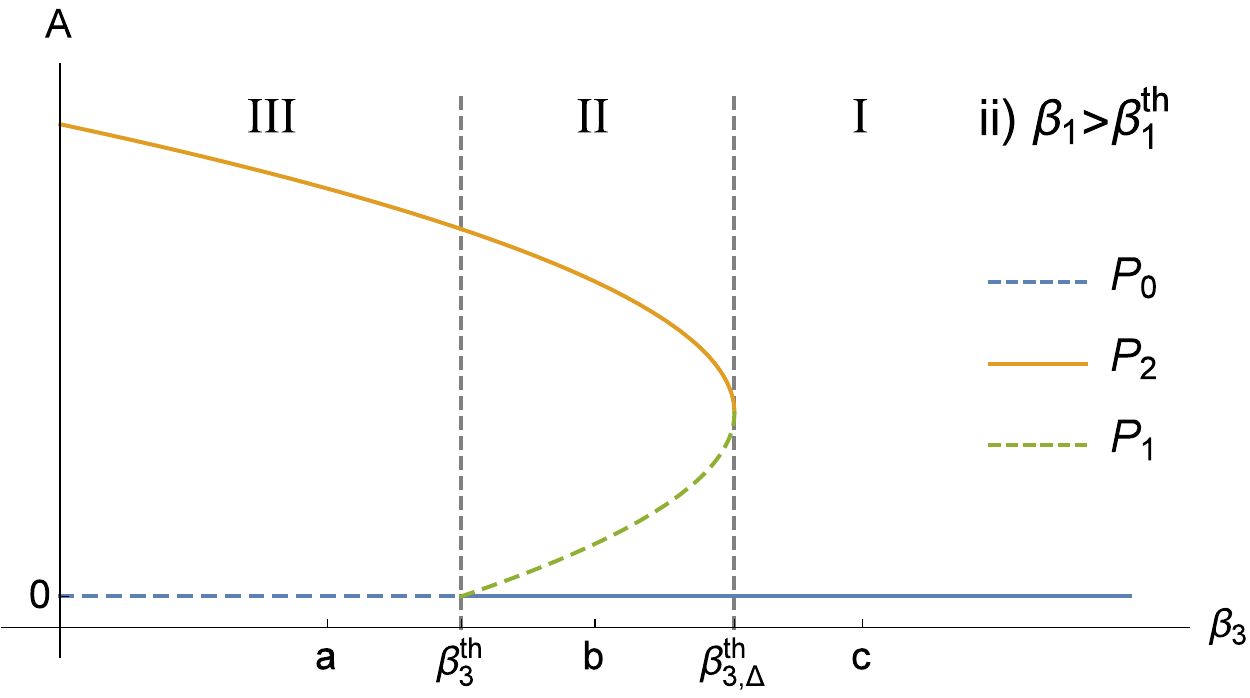}
\includegraphics[width=0.47\linewidth]{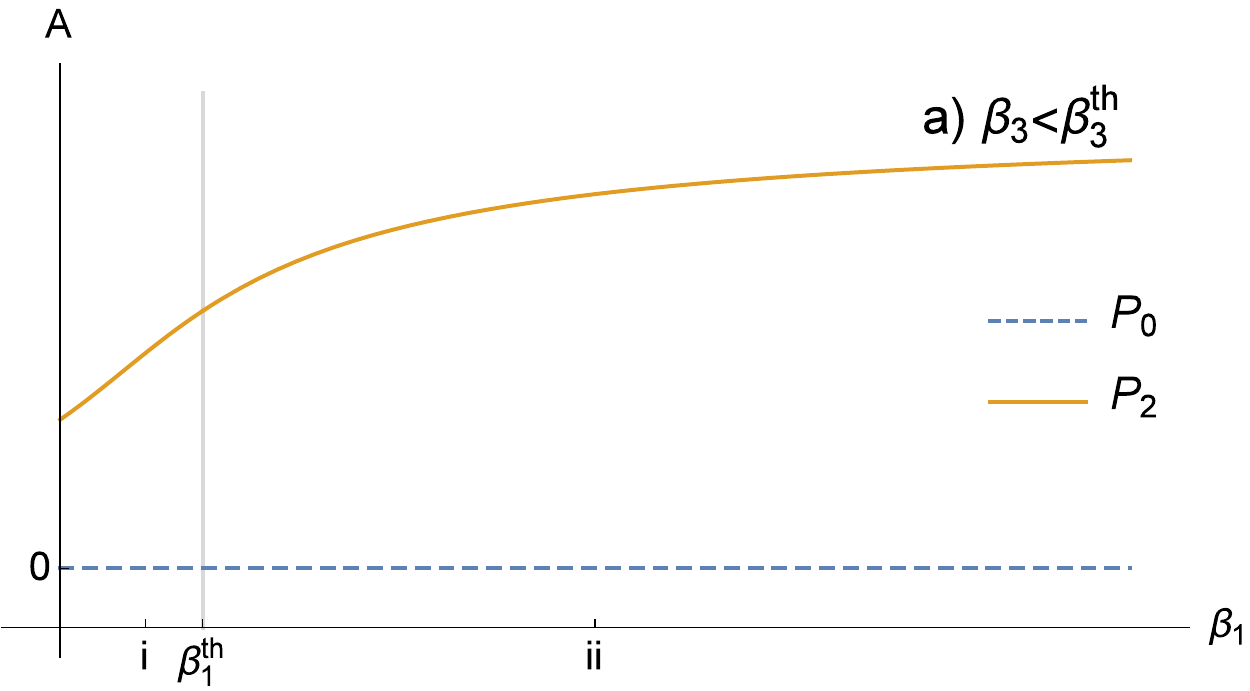}
\hfill
\includegraphics[width=0.47\linewidth]{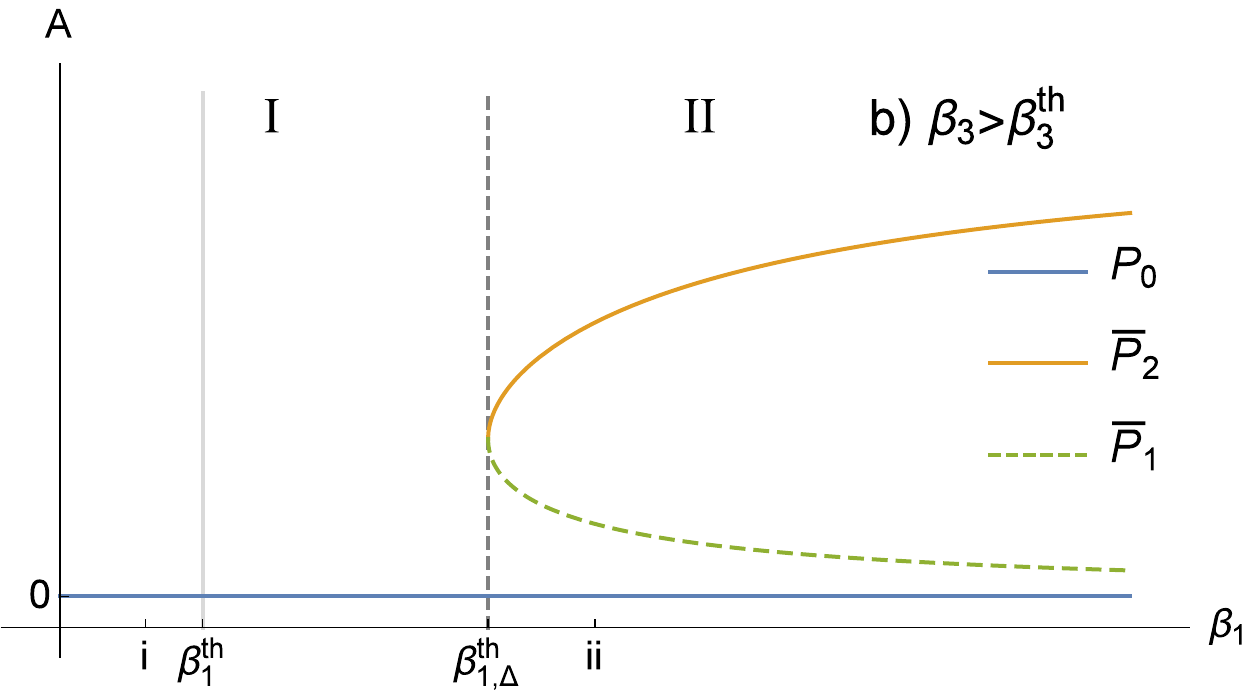}
\end{center}
\end{minipage}
\caption{Top: bifurcation diagrams of coordinate $A$ of equilibrium points when $\beta_3$ varies, with $\beta_1<\beta_1^{th}$ (left), and $\beta_1>\beta_1^{th}$ (right). Down: bifurcation diagrams when $\beta_1$ varies, with $\beta_3<\beta_3^{th}$ (left), and $\beta_3>\beta_3^{th}$ (right). Continuous plot corresponds to stable equilibrium, while dashed plot corresponds to unstable equilibrium. These diagrams are particular cases (horizontal sections i and ii and vertical sections a and b) of the bidimensional one, shown in Figure \ref{fig-bif-b1b3}.}
\label{fig-bifb3}
\end{figure}
\end{center}

We present in Figure \ref{fig-bifb3} bifurcation diagrams when $\beta_3$ and $\beta_1$ vary. When $\beta_3$ varies with $\beta_1$ fixed, two different cases occur (Figure \ref{fig-bifb3}, top panels). In the case i), the tumor is not much aggressive to normal cells, $\beta_1<\beta_1^{th}$, and we have a transition between regimes I and III through a transcritical bifurcation. In case ii), the tumor is very aggressive, $\beta_1>\beta_1^{th}$, and we have transitions from regimes I to II (saddle-node bifurcation) and from II to III (transcritical bifurcation). Before entering in regime III, there is a previous and additional interval, $\beta_3^{th}<\beta_3<\beta_{3,\Delta}^{th}$ that allows cancer onset depending on initial conditions (regime II). The comparison of cases (i) and (ii) contributes to the debate of whether intrinsic or extrinsic factors are the major responsible for cancer onset \cite{wu2016substantial,tomasetti2015variation}. Due to the possibility of a direct transition from regime I to II when $\beta_1>\beta_1^{th}$, we conclude that extrinsic factors appear to be the major cause of aggressive tumors, but combined with small contribution of intrinsic factors ($\beta_3<\beta_{3,\Delta}^{th}$). On the other hand, the possibility of direct transition from I to III in case (i) implies that non-aggressive tumors may arise due to intrinsic factors only.

When $\beta_1$ varies while $\beta_3$ is kept constant, we have a different effect (Figure \ref{fig-bifb3}, bottom panels). In the case when the tissue response to tumor is low, $\beta_3<\beta_3^{th}$, no bifurcation occurs and we remain in regime III for all $\beta_1$. In the second case, the tissue response is high, $\beta_3>\beta_3^{th}$, and a transition between regimes I and II occurs through a saddle-node bifurcation. We remain in regime II for all values of $\beta_1>\beta_{1,\Delta}^{th}$, and no transition to regime III occurs, on the contrary to case (b) when $\beta_3$ varies with a high aggressiveness of tumor. Comparison of diagrams (ii) and (b) lead to an interesting conclusion. While normal cells are able with their own characteristics (strong repair system) to guarantee tissue integrity against aggressive tumors, aggressive cancer cells, on the other hand, depend on genetic instability (initial conditions) when fighting aggressive normal cells.

\subsection{Cancer treatment as an attempt to allow a critical transition}

\begin{center}
\begin{figure}[!htb]
\begin{minipage}[b]{\linewidth}
\begin{center}
\includegraphics[width=.48\linewidth]{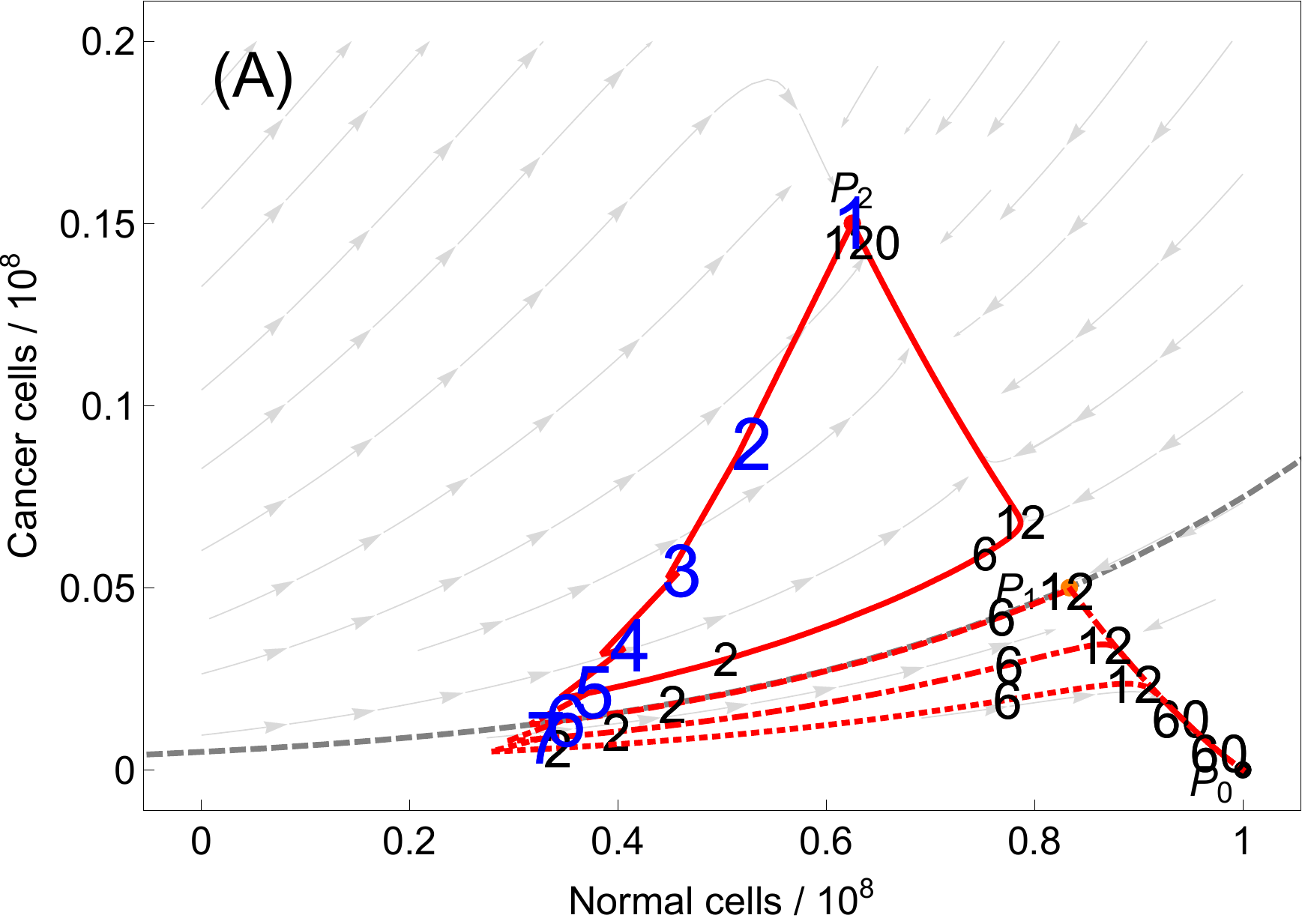}
\hfill
\includegraphics[width=.48\linewidth]{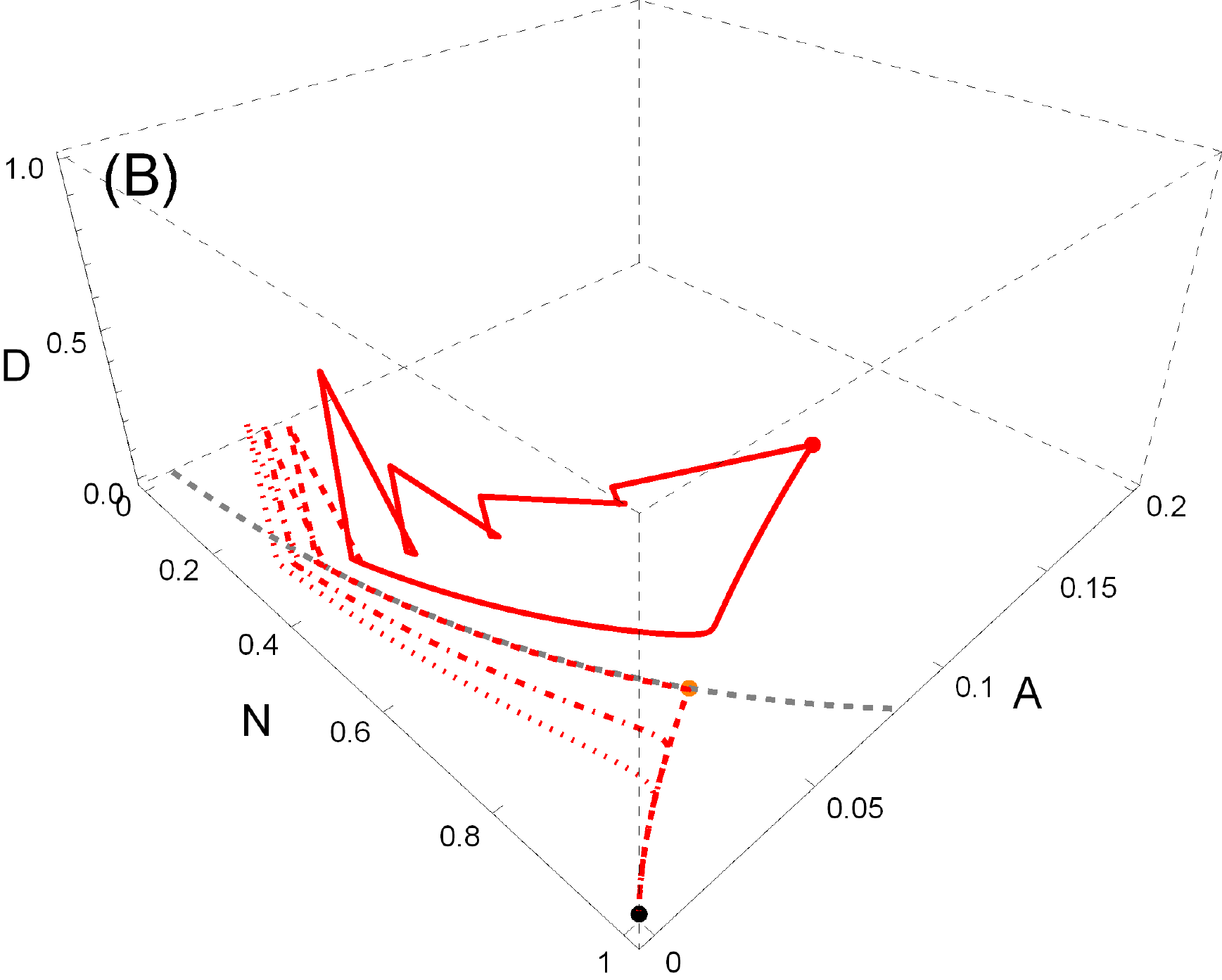}
\\
\includegraphics[width=.48\linewidth]{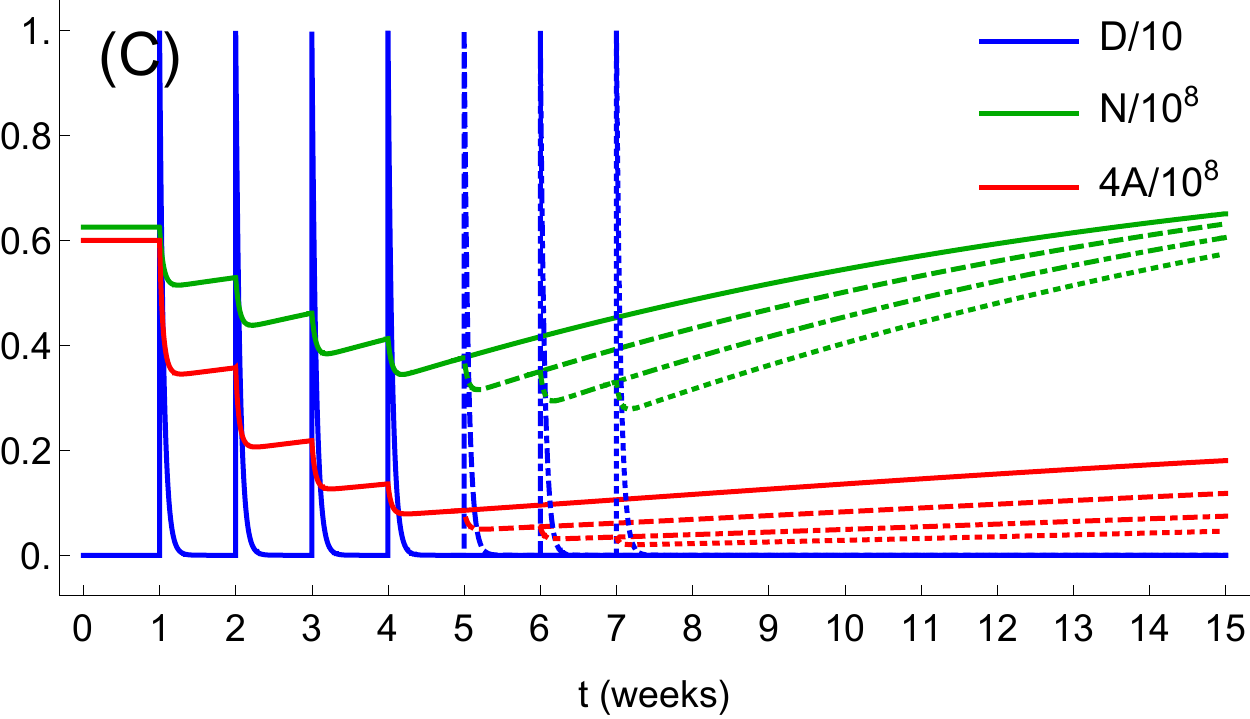}
\hfill
\includegraphics[width=.48\linewidth]{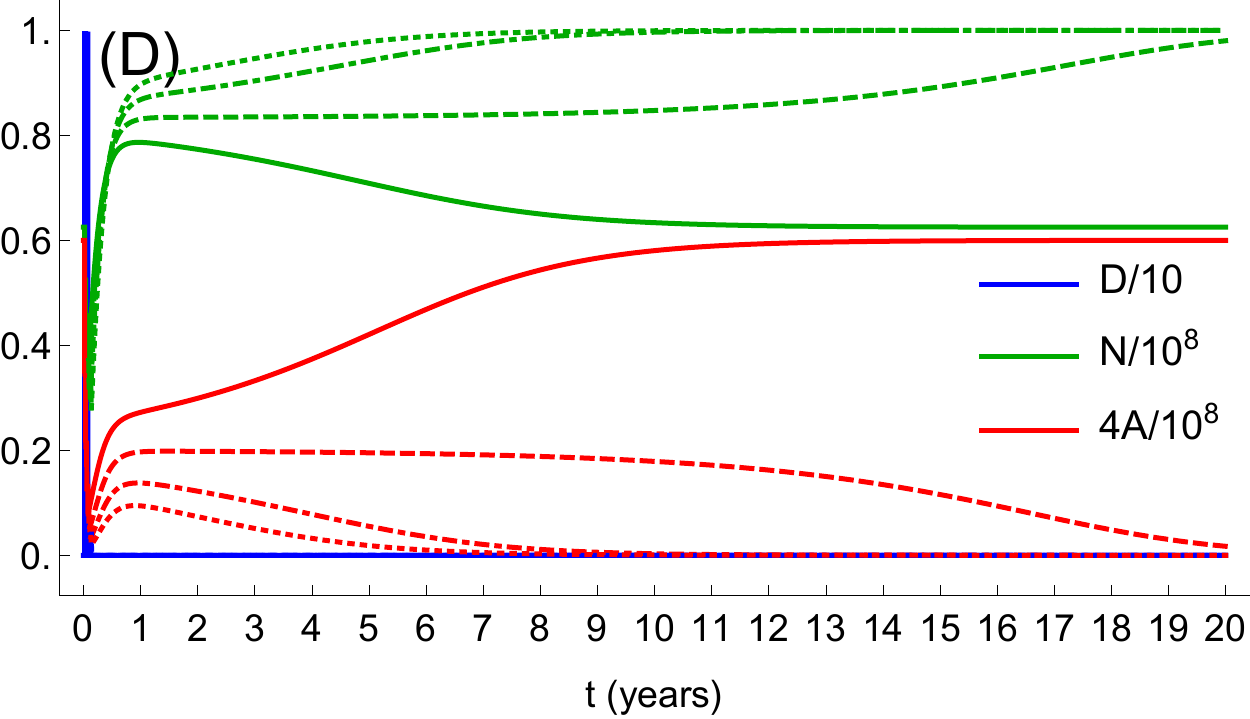}
\end{center}
\end{minipage}
\caption{Solutions of system \eqref{sisNAD_NA} when parameters correspond to regime II, with the number of doses given by $n=4$ (continuous), $n=5$ (dashed), $n=6$ (dot-dashed) and $n=7$ (dotted). Initial conditions were $(A(0),N(0))=P_2$, representing that the treatment was 
initiated only after the tumor reached the steady state $P_2$. In panel (A), the small black (large blue) numbers 
on phase portrait indicate the time in months (weeks) in which the solution was at each point. In panels (A) and (B) the gray dotted curve represents the separatrix between the basins of attraction of 
$P_0$ and $P_2$.}
\label{fig_NA_trat_II}
\end{figure}
\end{center}

\begin{center}
\begin{figure}[!htb]
\begin{minipage}[b]{\linewidth}
\begin{center}
\includegraphics[width=.48\linewidth]{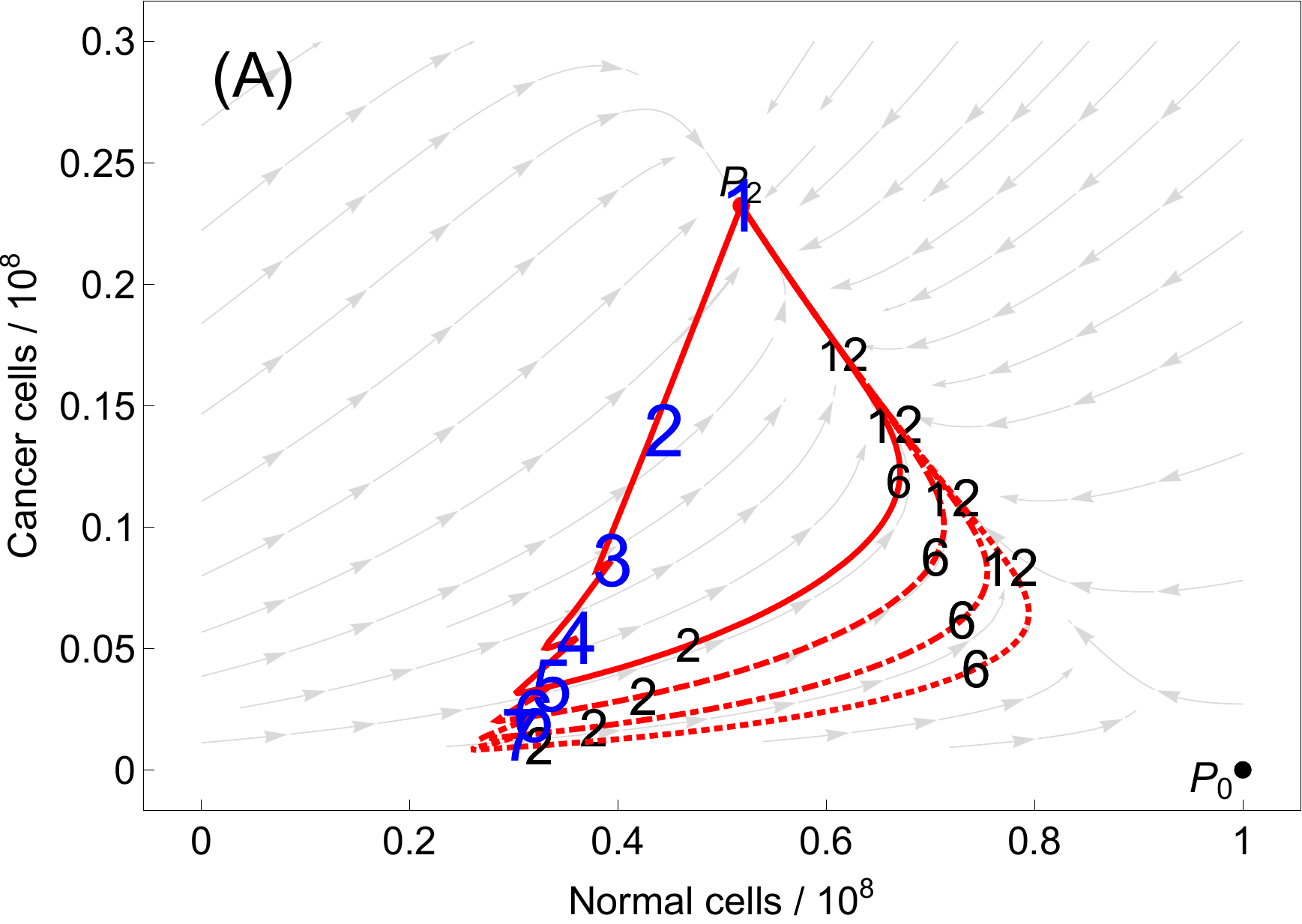}
\hfill
\includegraphics[width=.48\linewidth]{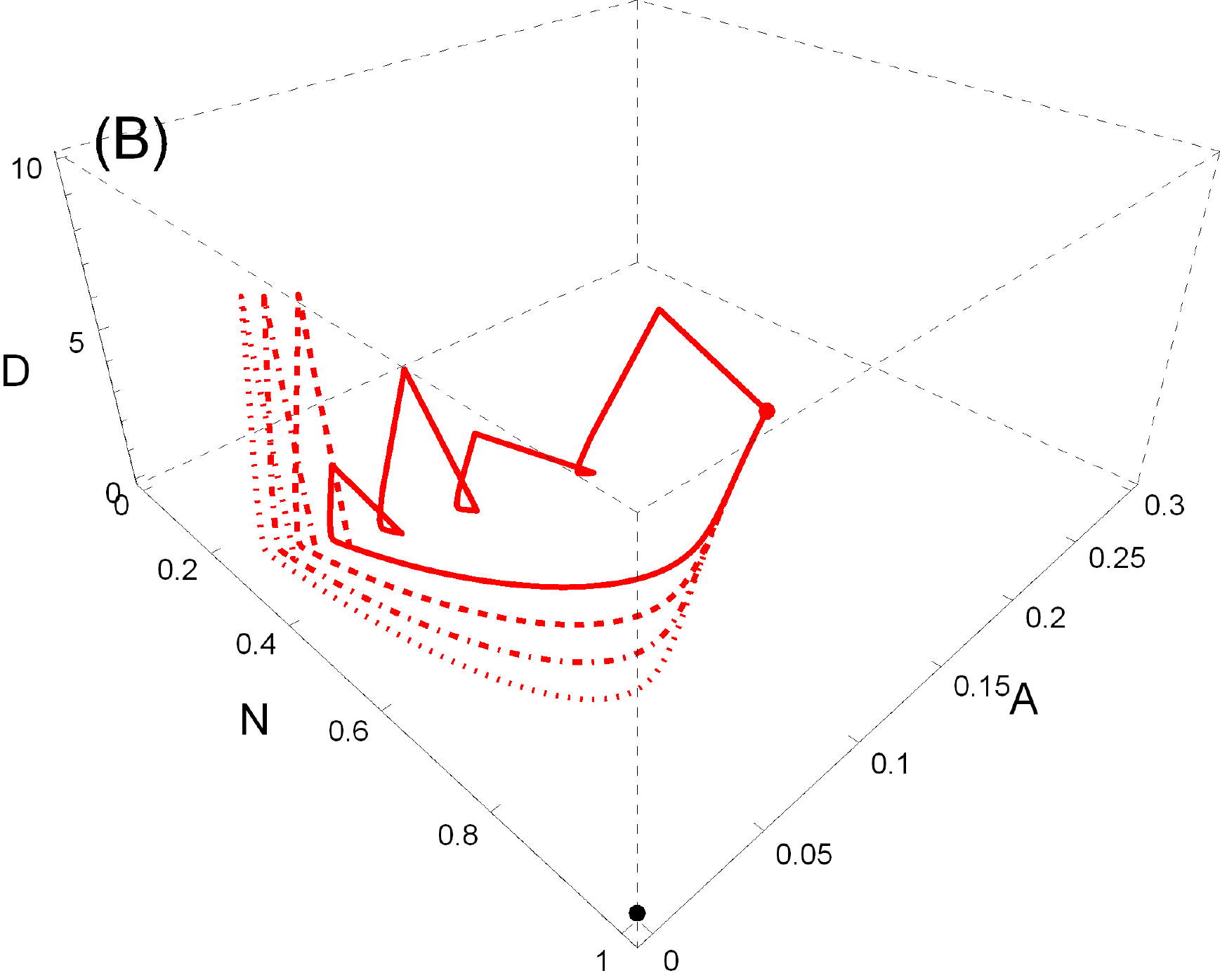}
\\
\includegraphics[width=.48\linewidth]{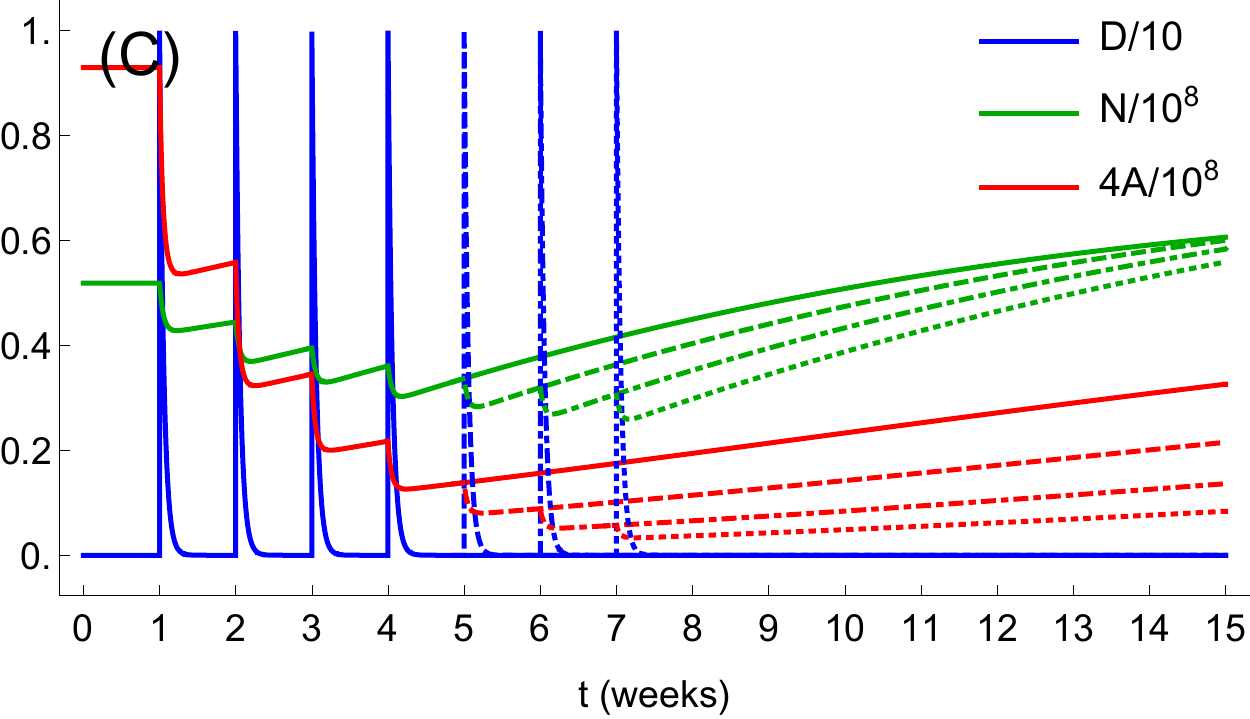}
\hfill
\includegraphics[width=.48\linewidth]{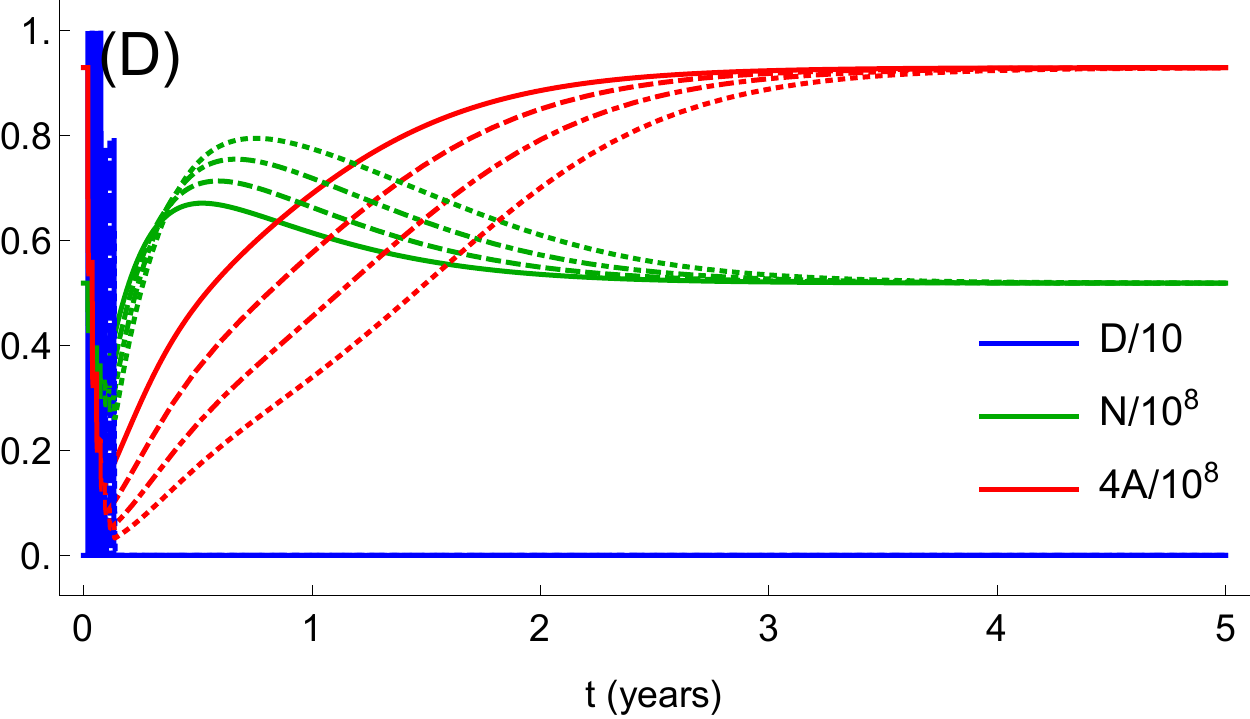}
\end{center}
\end{minipage}
\caption{Solutions of system \eqref{sisNAD_NA} when parameters correspond to regime III, with the number of doses given by $n=4$ (continuous), $n=5$ (dashed), $n=6$ (dot-dashed) and $n=7$ (dotted). Initial conditions were $(A(0),N(0))=P_2$, representing that the treatment was 
initiated only after the tumor reached the steady state $P_2$. In panel (A), the small black (large blue) numbers 
on phase portrait indicate the time in months (weeks) in which the solution was at each point.}
\label{fig_NA_trat_III}
\end{figure}
\end{center}

Now, let us consider cancer treatment. We focus on application of chemotherapy due to its wide use, but some of the general results may be extended to other types of treatment, like radiotherapy or surgery. A simple way to include chemotherapy in system \eqref{sisNA} is to consider the following equations:
\begin{subequations}
\label{sisNAD_NA}
\begin{align}
& & \dfrac{dN}{dt}=&\;r_N - \mu_N N - \beta_1 N A- \alpha_N \gamma_N ND
\label{dNdt2} \\
& & \dfrac{dA}{dt}=&\;r_A A \left(1-\frac{A}{K_A}\right)- (\mu_A+\epsilon_A) A  - \beta_3 N A - \alpha_A \gamma_A A D
\label{dAdt2} \\
& & \dfrac{dD}{dt}=&\;v(t)- \gamma_N ND- \gamma_A AD- \tau D.
\label{dDdt2}
\end{align}
\end{subequations}
Here, $D$ is the chemotherapeutic drug. It is administered according to a treatment function $v(t)$, and has a clearance rate $\tau$. The terms $\gamma_N ND$ and $\gamma_A AD$ describe drug absorption by normal and cancer cells, while the killing terms $\alpha_A \gamma_A AD $ and $\alpha_N \gamma_N ND$ follow the log-kill hypotheses. This is based on an analogy with the mass action law, and states that the exposure to a given amount of drug kills a constant fraction of the cell population  \cite{benzekry2015metronomic}. The treatment function can be described as a finite sum of Dirac Deltas, 
\begin{equation*}
v(t) = \displaystyle \sum_{i=1}^{n} \rho \delta (t-iT),
\end{equation*}
which represents $n$ doses of $\rho$ mg of drug each, each $T$ days. Figures \ref{fig_NA_trat_II} and \ref{fig_NA_trat_III} show simulations of system \eqref{sisNAD_NA} in cases II and III respectively. In these simulations, we assumed the values $\tau=2.5$ day$^{-1}$, $\gamma_A=0.3 \times 10^{-8}$ cell$^{-1}$day$^{-1}$, and $\alpha_A=0.5 \times 10^8$ cell mg$^{-1}$ and $\gamma_N=0.6\gamma_A$ and $\alpha_N=0.6\alpha_A$, since the chemotherapeutic agent is supposed to be more specific to cancer cells than normal cells. The values for $\gamma_A$ and $\alpha_A$ were taken arbitrarily, but with a order of magnitude coherent with other parameters and with the expected behavior for system \eqref{sisNAD_NA}. The treatment parameters were $\rho=10$ mg, $T=7 $ days, and $n=4,5,6,7$ doses.

More important than the form used to modeling the treatment $v(t)$, the general property is that all chemotherapeutic treatments cease after some time $t_f>0$, i.e., $v(t) =0$ for all $t>t_f$. Thus, $D(t)\leq -\tau D $  and so $D \to 0$ when $t \to \infty$. Therefore \textit{all solutions of system \eqref{sisNAD_NA} approach solutions of system \eqref{sisNA}} when $t \to \infty$. This fact have an important consequence in our ecological perspective. In the real system underlying tumor growth, the treatment would have only the effect of state space disturbance, without altering the intrinsic dynamics. This important feature reveals that the possibility of cure, above all, concerns questions of stability and resilience. Figures \ref{fig_NA_trat_II} and \ref{fig_NA_trat_III} provide an illustration of this fact.

If tumor growth in a patient is described by some underlying dynamical system which does not have a cancer cure stable equilibrium, then a complete cure is not possible. This is what happens in case III, due to the weakness of the repair system, and it is illustrated in Figure \ref{fig_NA_trat_III}. After getting near this equilibrium through the treatment, the system moves back to the cancer equilibrium, even if it takes a long time. Thus, it is necessary a systemic change that alters the dynamics permanently. However, as also shown in Figure \ref{fig_NA_trat_III}, even in this regime of instability of the cure equilibrium, treatment may lead to large time survival. Indeed, the closer the system approaches the cure equilibrium, the longer it takes to tumor recurrence be observable. This fact is related with the large time necessary to pass through a saddle-point.

On the other hand, even if the system has a cancer cure stable equilibrium, as in case II here, the treatment may be ineffective if the solution does not achieve the basin of attraction of the cure equilibrium. It is the case of simulation with $n=4$ in case II, shown in Figure \ref{fig_NA_trat_II}. In other words: a necessary condition to a treatment be effective is that the underlying system must have a stable cancer cure equilibrium; and the sufficient condition for the treatment be effective is that the treatment must move the trajectory to the basin of attraction of this equilibrium (simulations with $n=5,6,7$ in case II). Once it has been reached, the treatment can stop, since the patient own repair system will eliminate the reminiscent cancer cells, and move the trajectory in direction to the cure equilibrium. Thus, the resilience of cancer equilibrium plays an important role: if this equilibrium has a large and deep basin, and is located at a large distance from the basin boundary, then more doses, or more intense doses, will be necessary in order to make the treatment to be effective. Further, it also suggests a mechanism through which two individuals with similar diagnosis and treatments may have different fates: some of those treatments which end very near the separatrix (simulation with $n=5$ in Figure \ref{fig_NA_trat_II}) may become ineffective due to stochastic fluctuations which can drive the system to the cancer basin again, while other may continue in the cure basin. This indicates that truly effective treatments should drive the system to a safe distance from the boundary of the basin. This rationale agrees with the fact that treatments which consist of single surgery or radiotherapy (which would correspond to large state space disturbances) must be reinforced by subsequent adjuvant treatment in order to preclude tumor relapse \cite{stupp2005radiotherapy}.

\section{Resilience Analysis}
\label{secNA:res}

\begin{center}
\begin{figure}[htb!]
\begin{minipage}[b]{\linewidth}
\begin{center}
\includegraphics[width=.9\linewidth]{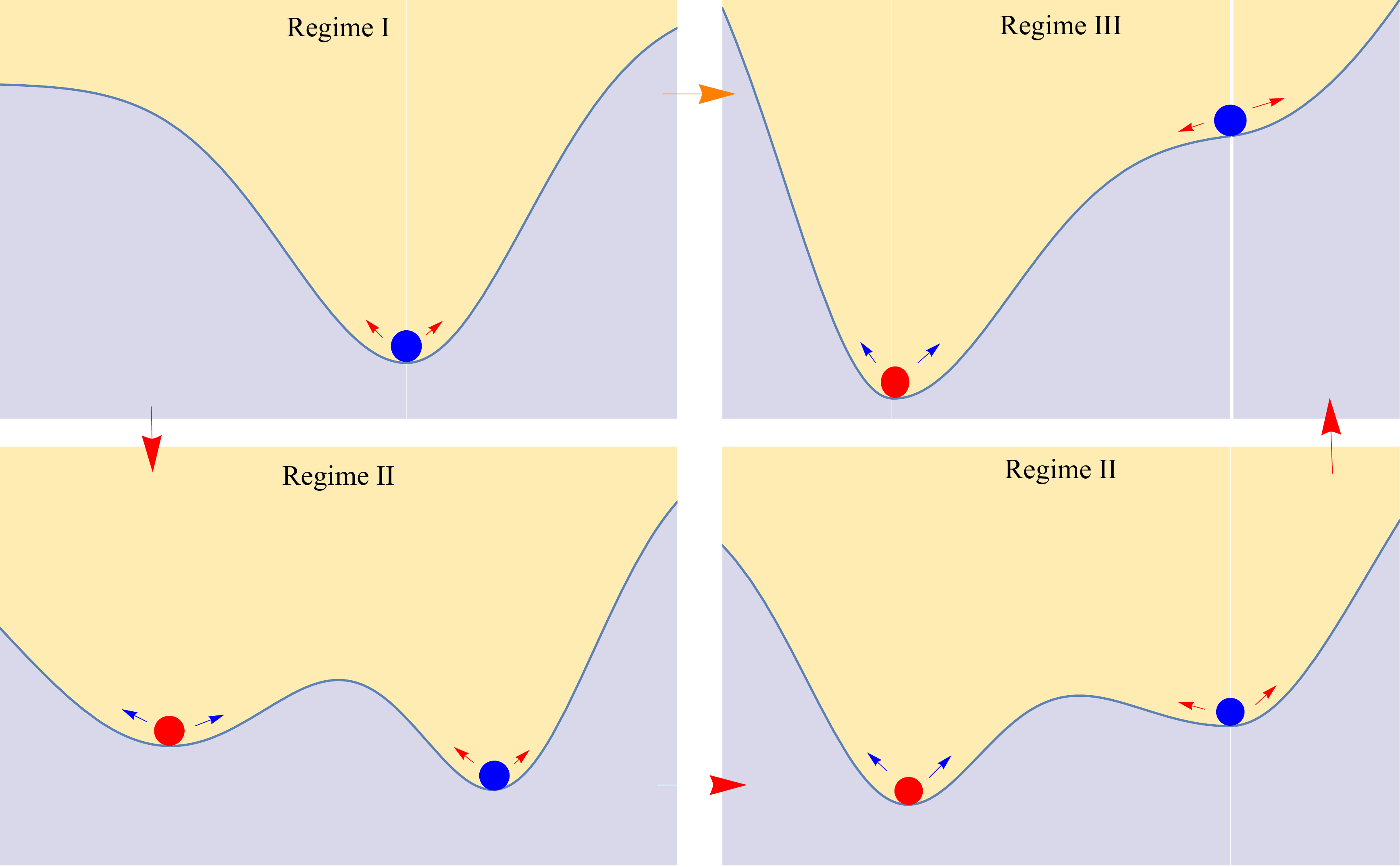}
\end{center}
\end{minipage}
\caption{An ecological view on cancer onset and treatment as the switching between two states, a cancer cure state (positions of blue balls) and a cancer state (positions of red balls). Perturbations on state variables are due to exposure to external carcinogenic factors and genetic instability (small red arrows), which favor the moving in the direction to the cancer state, or to chemotherapy, radiotherapy and surgery (small blue arrows), which move the system towards the cancer cure equilibrium. Perturbations in parameters (large arrows) lead to transitions between the three different regimes (I, II and III), and may create or destroy equilibria, or lead to changes of stability in these equilibria, what can makes impossible either the cure or the onset of cancer.}
\label{fig_NA_terreno}
\end{figure}
\end{center}

The ecological point of view on cancer discussed above is illustrated through the stability landscape in Figure \ref{fig_NA_terreno}. In this stability landscape, the creation of the cancer equilibrium and the loss of stability by the cancer cure equilibrium (large red and orange arrows) is achieved by the sequential acquiring of genetic alterations that improve the fitness of cancer cells (the hallmarks of cancer \cite{hanahan2011hallmarks}), both by deregulating mechanisms of control or by creating mechanisms which favor cancer cell functions. The transition from I to III (large orange arrow) occurs only if tumor aggressiveness is low. Destruction of cancer equilibrium and creation of stability of the cancer cure equilibrium (opposite directions of large red and orange arrows) will be achieved only by a permanent change in the intrinsic system, through, for instance, restoring the control systems (immunostimulation for example) or by limiting cancer cells functions (continuous anti-angiogenic treatments for example, which would decrease the value of $K_A$ in theory). Finally, combination of small changes in parameters with perturbations on state variables also have a fundamental role, since the former can diminish the basin of attraction while the latter push the system to cross the basin boundary.
 
Let us analyze the effect of parameters changes on the resilience of stable equilibria of system \eqref{sisNA}, i.e., on size and shape of their basins of attraction. To perform this analysis we briefly review the measurements which may be of importance when analyzing the resilience of an equilibrium \cite{mitra2015integrative,meyer2016mathematical,walker2004resilience,menck2013basin}. As far as we know, very few papers have dealt with this kind of analysis for population dynamics models \cite{mitra2015integrative,fassoni2014basins,fassoni2014mathematical}, although it is not an uncommon approach for systems modeling power grids \cite{menck2013basin}. Recently, Mitra \textit{et al.} applied these measures to the nonlinear pendulum, the daisy-world model, and to an one-dimensional model of desertification in Amazon forest \cite{mitra2015integrative}. Also, as far as we know, the methods provided here to calculate these measures are novel and more efficient than the currently used ones, which consist in integrating the system in many points of the phase space, assigning each point to some basin of attraction. Throughout this section, we represent a point with coordinates $N$ and $A$ by $X=(N,A)$.

\subsection{The latitude $L(P)$ of an equilibrium point}
The \textit{latitude} of an equilibrium point corresponds to the volume of the basin of attraction. It measures the resilience of that equilibrium with respect to state-space perturbations. The larger the latitude of an equilibrium, the smaller is the chance that external or probabilistic events will drive the system outside the basin of attraction. For two dimensional systems, the latitude corresponds to the area of the basin of attraction. However, as it may happen that the basin of attraction has infinite area, it may be the case to consider its area inside a relevant bounded region. In the case of system \eqref{sisNA}, all trajectories remain in the box $B$ given in \eqref{boxB}. This box could be this bounded region of interest. However, biological relevant perturbations of $P_2=(N_2,A_2)$ will diminish the value of both coordinates, and relevant perturbations of $P_0=(r_N/\mu_N,0)$ will diminish the first coordinate and increase the second one. Thus, we consider the smallest box $C$ which contains $P_0$ and $P_2$ as our bounded region of interest. It is clear that $$C=[0,r_N/\mu_N]\times [0,A_2].$$ Thus, if ${\cal A}(P)$ denotes the basin of attraction of $P$, then the latitude of $P$ in our case is defined by
\begin{equation}
L(P) = \dfrac{ \textrm{ Area}\, \left( {\cal A}(P) \cap C\right) }{\textrm{ Area}\, \left( C\right)}.
\label{latP}
\end{equation}
We divide by the total area of $C$ to obtain a non-dimensional quantity normalized between $0$ and $1$. In the case when system presents bistability (region II of parameters space) it is clear that
\[
L(P_0)+L(P_2)=1.
\]
Extreme cases are regions I and III. In region I, when $P_0$ is globally stable, we have
\[
L(P_0)=1, \ L(P_2)=0,
\]
In region III, when $P_2$ is globally stable,
\[
L(P_0)=0, \ L(P_2)=1.
\]

\label{appCalcL}

We present a simple and efficient method to calculate $L(P)$ for two dimensional systems. It consists of two steps. The first step is to obtain the terms of a series expansion for a parametric representation for the stable manifold of saddle point in a two-dimensional system. The second step involves the use of Green's Theorem to transform the area of the basin of attraction in a line integral calculated along the stable manifold approximated in the first step.

\paragraph{Summary of first step}
Let $X^*$ be a saddle point of the two-dimensional system
\[
X'=F(X), \ X \in \mathbb{R}^2,
\]
where $F$ is a $C^1$ vector field. In the vicinity of $X^*$, this system can be re-written as
\[
X'=J(X-X^*)+G(X),
\]
where $G(X)={\cal O}(||X-X^*||^2)$ and $J=F'(X^*)$ is the Jacobian matrix evaluated at $X^*$. Let $J=MKM^{-1}$ its Jordan decomposition, with
\[
K=
\left[
\begin{array}{cc}
k_{11} & 0 \\
0 & k_{22}
\end{array}
\right],
\]
where $k_{11}$ and $k_{22}$ are the eigenvalues of $J$ and satisfy $k_{11}<0<k_{22}$, since $X^*$ is a saddle-point. With the change of coordinates
\[
U=M^{-1}(X-X^*),
\]
the previous system becomes
\[
U'=KU+R(U),
\]
where $R(U)=M^{-1}G(X^*+MU)={\cal O}(||U||^2)$. This system can be written in coordinates $U=(u,v)$ as
\[
u'=f(u,v),\ \
v'=g(u,v).
\]
The origin is a saddle point for this system. Its stable manifold is tangent to $v=0$, and is locally the graph of a function $v=p(u)$. Substituting it into $v'=g(u,v)$ we obtain a nonlinear ODE for $p(u)$:
\[
p'(u)f(u,p(u))=g(u,p(u)).
\]
Although this ODE cannot be solved in general, we can write a series expansion
\[
p(u)=c_2u^2+c_3u^3+\cdots
\]
and solve term by term, obtaining the coefficients $c_j$ recursively. For large $k$, the truncated polynomial $p_k(u)=c_2u^2+\cdots+c_ku^k$ provides a good approximation 
 manifold near the origin. Returning to the original coordinates, $$s_k(u)=X^*+M\left(u,p_k(u)\right)^T$$ is a parametric approximation of the stable manifold of $X^*$, for small $||u||$.

\paragraph*{Summary of the second step}
Now we explain how to calculate
\begin{equation}
L(P) = \dfrac{ \textrm{ Area}\, \left( {\cal A}(P) \cap C\right) }{\textrm{ Area}\, \left( C\right)} = \dfrac{\iint_{{\cal A}(P) \cap C  }dAdN}{\iint_{ C} dAdN}.
\label{latPapp}
\end{equation}
The infinitesimal area element is denoted by $dAdN$. The integral in the denominator is easy to calculate in general; in our case $C$ is a rectangle. The difficulty lies in calculating the integral in the numerator. However, it can be easily calculated by using the approximation of the stable manifold obtained in the previous step and Green's Theorem. Indeed, the boundary of the region ${\cal R}= {\cal A}(P) \cap C$ is formed by four curves, $\partial {\cal R}=S\cup L_1 \cup L_2  \cup L_3$, where $S$ is the part of the stable manifold of $P_1$ which is contained in $C$, and $L_i$, $i=1,2,3$, are line segments contained in the boundary of $C$. By Greens' Theorem, the integral can be written as
\[
\iint_{{\cal A}(P) \cap C }dAdN=\oint_{\partial {\cal R} } N dA=\int_{ S } N dA+\sum_{i=1}^3\int_{ L_i } N dA,
\]
with the correct orientation defined for these curves. The summation terms are easily computed, and a good approximation for the integral in $S$ is given by using the parametric approximation obtained in the first step. With this approach, we have a very good approximation of $L(P)$.

Let us comment on the applicability of this method. First of all, the method works when the separatrix is formed by invariant manifolds of saddle points \cite{chiang1988stability}. If many saddle points lie in the boundary, then the first step needs to be applied to each point. Second, the bounded region $C$ must be such that its area is easily calculated. These first two requirements are very general \cite{chiang1988stability}. The last and most restrictive condition is that, for each saddle point in the separatrix, the local approximation obtained in the first step must be a good approximation at all points inside $C$. The method fails when this requirement is not satisfied. In our case, for parameters values in Table \ref{tablePar}, the approximation for the stable manifold of $P_1$ with 25 terms was obtained in a few seconds by \textit{Mathematica}$^\copyright$, and provided a very good approximation for points inside $C$. The separatrix in Figures \ref{fig_NA_onset}, \ref{fig_NA_trat_II} and \ref{fig_NA_ilustra_met} was plotted with this approximation.

\subsection{The precariousness $Pr(P)$ of an equilibrium point}
\begin{center}
\begin{figure}[!htb]
\begin{minipage}[b]{\linewidth}
\begin{center}
\includegraphics[width=.75\linewidth]{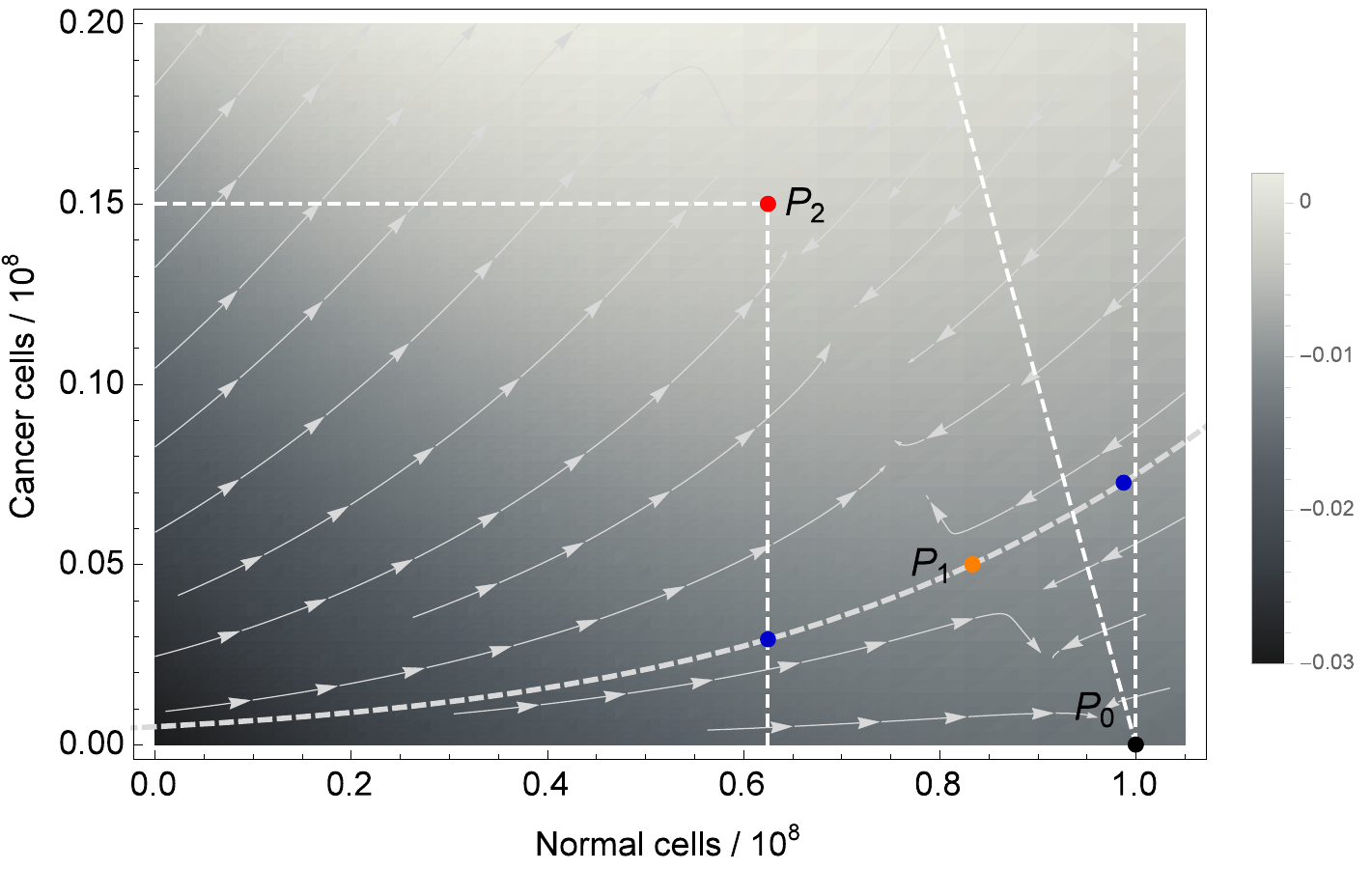}
\end{center}
\end{minipage}
\caption{Phase portrait of system \eqref{sisNA}. $Z(P_2)$ is the box with $P_2$ in one of the vertices. $Z(P_0)$ is the triangular region with $P_0$ in its vertices. The blue points are the nearest points of $P_0$ and $P_2$ inside regions $Z(P_0)$ and $Z(P_2)$ respectively. The intensity of color on background correspond to the value of the local resistance $R(N,A)$ defined in sub-section \ref{subsec:Resist}.
}
\label{fig_NA_ilustra_met}
\end{figure}
\end{center}

Another important measure is the \textit{precariousness} of an equilibrium point. Roughly speaking, it is defined as the minimum perturbation required to drive the system to another basin of attraction. As a basin of attraction of interest may be large but the equilibrium may be located near the boundary, the precariousness is an important measure. It can be defined as
\[
Pr(P) = \inf \{ \text{dist} (P,X),\ X \in \partial {\cal A}(P) \},
\]
where $\partial \Omega$ stands for the boundary of the set $\Omega$, and $\text{dist}(X,Y)$ is the Euclidean distance. However, it may happen that perturbations of relevance may not occur in the direction where this minimum distance is achieved. Thus, one can consider 
\[
Pr(P) = \inf \{ \text{dist}(P,X),\ X \in \partial {\cal A}(P)\cap Z(P) \},
\]
where $Z(P)$ is a specified set containing the relevant directions for perturbations from $P$ \cite{meyer2016mathematical}. For system \eqref{sisNA}, all relevant perturbations of $P_2$ will diminish both coordinates. Thus we define 
\begin{equation}
Pr(P_2) = \inf \{ \text{dist}(P_2,X),\ X \in \partial {\cal A}(P_2)\cap Z(P_2) \}, \ \ Z(P_2)=[0,N_2]\times [0,A_2].
\label{prP2}
\end{equation}
Let us now consider which are the relevant perturbations of $P_0$. As new cancer cells may arise in the tissue, we may consider perturbations which increase the value of $A$. Thus, it would be the case to consider a single direction, $\vec{v}_1=(0,1)$. However, as cancer cells arise during mitosis, roughly speaking, the number of normal cells diminish by one when a cancer cell arises. Thus, it would be the case to consider the direction given by $\vec{v}_2=(-1,1).$ Therefore, we consider all directions between $\vec{v}_1$ and $\vec{v}_2$, giving us the set above the line $ N+A = r_N/\mu_N$ (which passes through $P_0$ and is parallel to $\vec{v}_2$) and to the left of the vertical line $N=r_N/\mu_N$ (which passes through $P_0$ and is parallel to $\vec{v}_1$). Thus, the expression of $Z(P_0)$ is
\[
Z(P_0) = \{(N,A) \in \mathbb{R}^2_+, \ N+A\geq r_N/\mu_N, N\leq r_N/\mu_N\}.
\]
An illustration of regions $Z(P_0)$ and $Z(P_2)$ can be seen in Figure \ref{fig_NA_ilustra_met}.

In extreme cases I and III we do not have bistability, and $Pr(P)$ must be defined appropriately. In case III, when $P_2$ is globally stable, we define $Pr(P_2)$ as the tumor volume, $Pr(P_2)=A_2$, meaning that the removal of the entire tumor leads the system outside the basin of attraction of $P_2$. Indeed, this removal leads the system to the $N$-axis, which is the invariant manifold of $P_0$. Analogously, in case I, when $P_0$ is globally stable, we define $Pr(P_0)=N_0$. However, in this case the $A$-axis is not invariant, and another choice would be $Pr(P_0)=\infty$.

\label{appCalcPr}

The result obtained in the first step of the previous section can be used to calculate the precariousness of an equilibrium straightforwardly. In our case, let $s_k(u)=(s_k^{1}(u),s_k^{2}(u))$ be the approximation for the stable manifold of $P_1$ obtained in that step. To calculate $Pr(P_0)$, for instance, we first find the interval $I_0$ of values of $u$ such that $s_k(u)\in Z(P_0)$. The solution $u_1$ of equation
\[
s_k^{(1)}(u_1)=r_N/\mu_N
\]
is the value of $u$ such that $s_k(u)$ lies in the vertical line $N=r_N/\mu_N $. The solution $u_2$ of
\[
s_k^{(1)}(u_2)+s_k^{(2)}(u_2)=r_N/\mu_N
\]
is the value of $u$ such that $s_k(u)$ lies in the line $N+A=r_N/\mu_N $. Thus, $u_1$ and $u_2$ are the extrema of interval $I_0$. With this, $Pr(P_0)$ is obtained by solving the minimization problem
\[
\begin{array}{rc}
\min  & || s_n(u)-P_0||\\
 & u\in I_0
\end{array}
\]
which can be easily solved. The precariousness of $P_2$ is calculated in an analogous way.

\subsection{The resistance $R(P)$ of an equilibrium point}
\label{subsec:Resist}

Finally, the third important measure concerning resilience of an equilibrium is termed as the \textit{resistance} of this equilibrium. It refers to ``the ease or difficulty of changing the system, related to the topology of the
basin - deep basins of attraction indicate that greater forces or perturbations are required to change
the current state of the system away from the attractor'' \cite{walker2004resilience}. Thus, a resistant system will overcome perturbations rapidly, while a system with small resistance can be driven to a basin transition through a series of small perturbations that are not absorbed enough. This description concerns exactly the illustration provided by Figure \ref{fig_NA_trat_II}. There, if $P_2$ is more resistant, more chemotherapeutic doses, or more intense doses, would be necessary to drive the system to the basin of attraction of $P_0$.

The task of characterizing the sizes and frequencies of perturbations that a basin of attraction can absorb is currently a research area \cite{meyer2016mathematical}. Recently, Mitra and others \cite{mitra2015integrative} proposed an approach to measure the resistance of a point in state-space to local pertubations using local Lyapunov exponents \cite{abarbanel1991variation}. Again, consider the two-dimensional system
\[
X'=F(X),
\]
where $F$ is a $C^1$ vector field. The \textit{instantaneous Jacobian matrix at} $X$ is defined as
\[
{\cal J}_{dt}(X) = I_{2 \times 2} + J(X) dt,
\]
where $J(X)$ is the Jacobian at $X$ and $dt$ is an infinitesimal time. Let $\sigma_i(X)$, $i=1,2$, be the square roots of the eigenvalues of the right Cauchy-Green tensor ${\cal J}_{dt}(X)^T {\cal J}_{dt}(X)$. The $\sigma_i(X)$'s are also the singular values of ${\cal J}_{dt}(X)$. They measure the instantaneous stretching of the neighborhood of the trajectory at $X$. The local Lyapunov exponents evaluated at the state $X$ are defined as
\[
\lambda_i(X)=\dfrac{1}{dt}\ln \left(\sigma_i(X)\right),\ i=1,2,
\]
and measure the rate of stretching at $X$. The local resistance $R(N,A)$ at the state $X=(N,A)$ may be defined as
\begin{equation}
\label{locRes}
R(X)=-\max \{ \lambda_1(X),\,\lambda_2(X) \}.
\end{equation}
Thus, the resistance of an equilibrium point can be measured as
\begin{equation}
\label{resP}
Res(P)= \dfrac{\iint_{{\cal A}(P)\cap C} R(N,A) \,dNdA}{ \iint_{C} R(N,A) \,dNdA}.
\end{equation}
The division by the total resistance on $C$ is made to obtain a non-dimensional quantity normalized between $0$ and $1$.

\label{appCalcRes}

For system \eqref{sisNA}, analytical expressions for $\lambda_i(X)$, $i=1,2$, at state $X=(N,A)$ can be calculated using \textit{Mathematica}$^\copyright$. The density plot of $R(N,A)$ is showed in Figure \ref{fig_NA_ilustra_met}. With these expressions, we can calculate integrals in \eqref{resP}. The unique concern is with respect to the region of integration in the first integral, ${\cal A}(P)\cap C$. However, this integral can be calculated using the approximations for the stable manifold of $P_1$ obtained in the first step of Section \ref{appCalcL}.

\subsection{Application of resilience analysis to system \eqref{sisNA}}

\begin{center}
\begin{figure}[!htb]
\begin{minipage}[b]{\linewidth}
\begin{center}
\includegraphics[width=.6\linewidth]{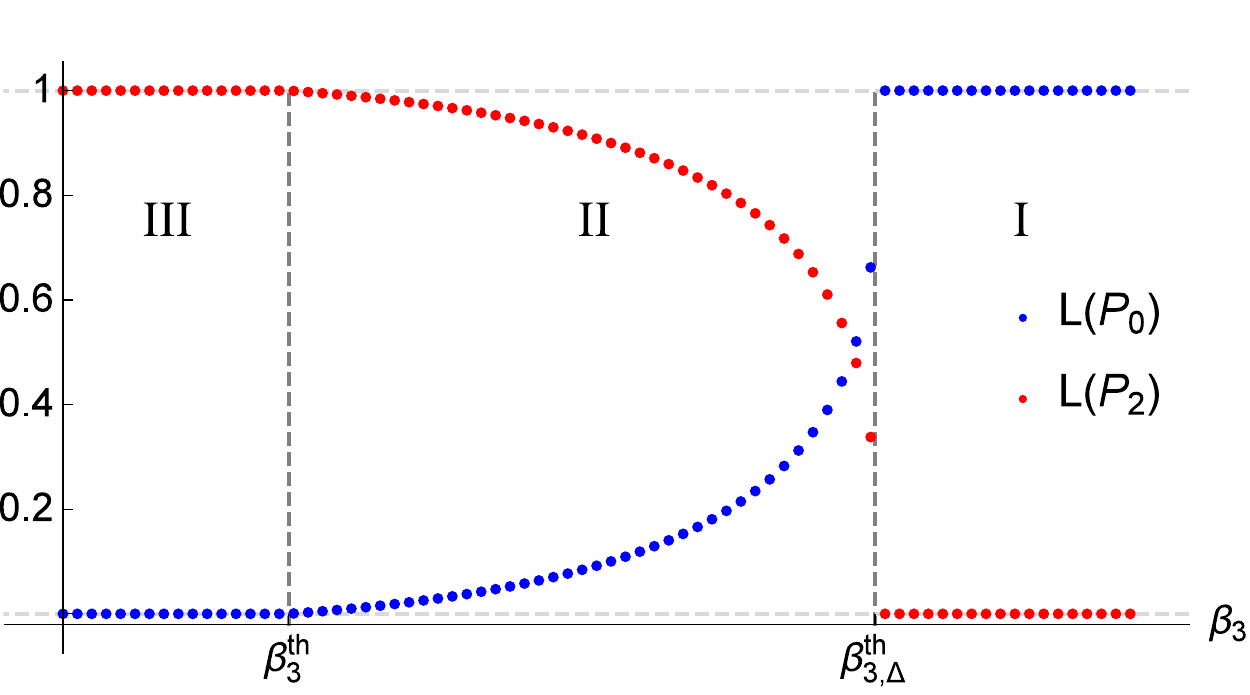}
\includegraphics[width=.6\linewidth]{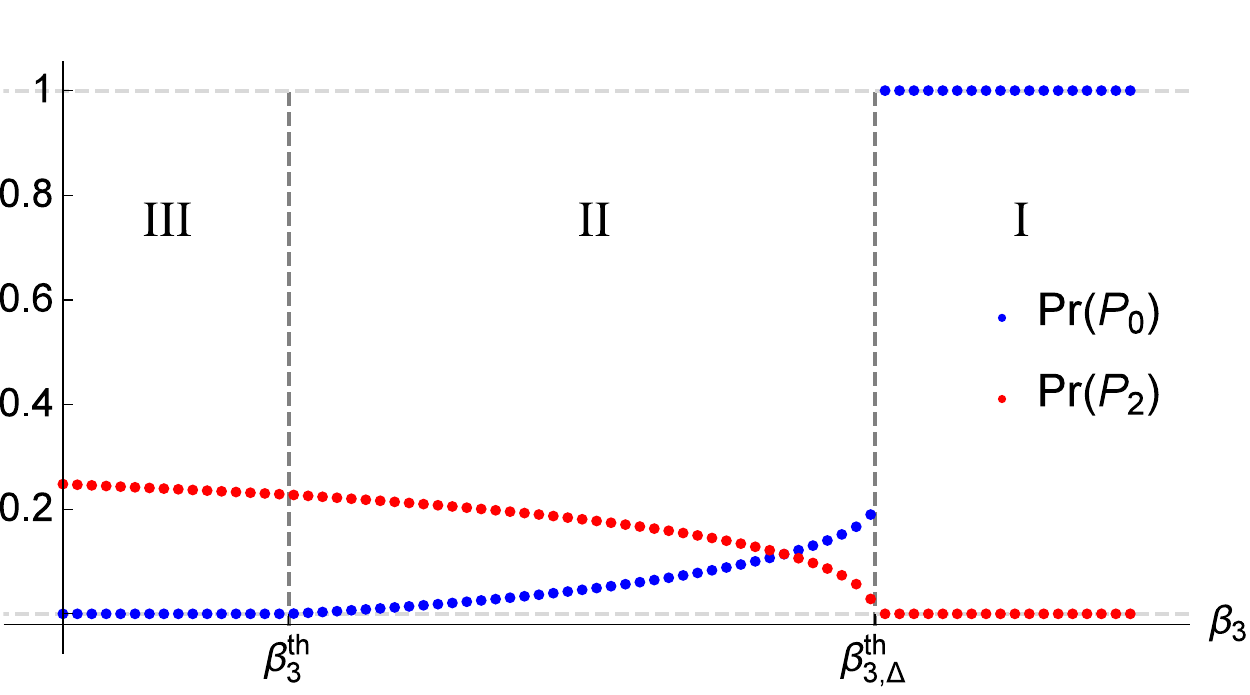}
\includegraphics[width=.6\linewidth]{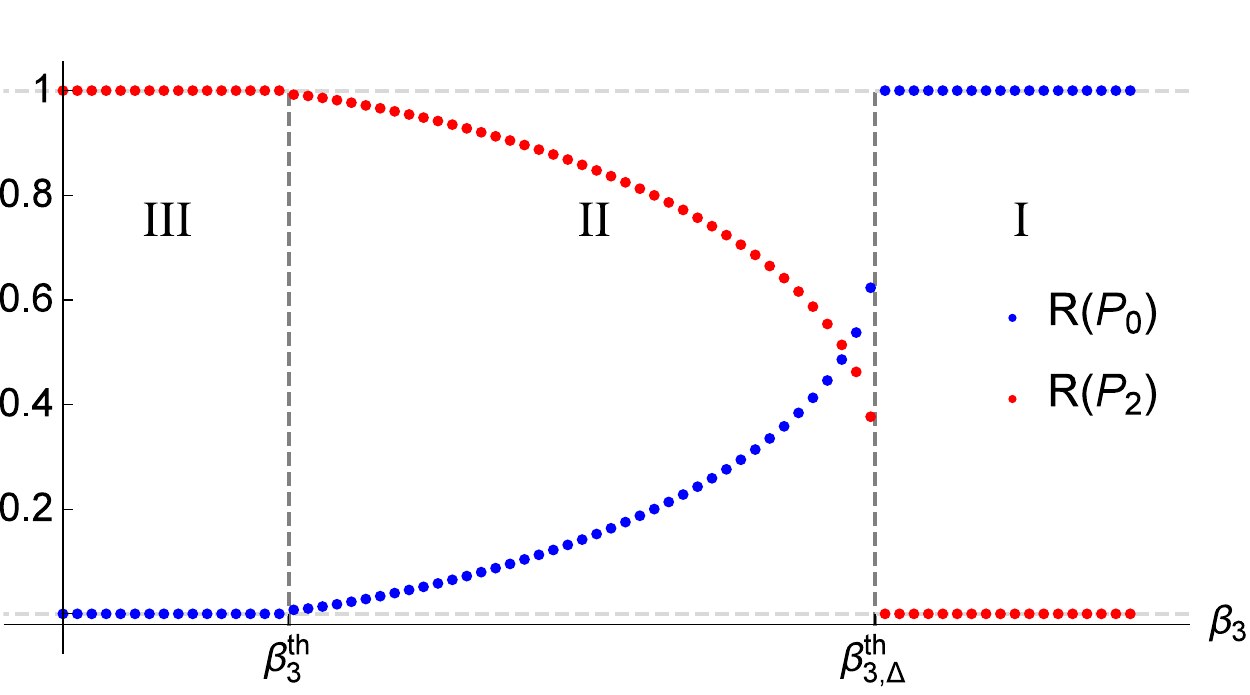}
\end{center}
\end{minipage}
\caption{Values of $L(P_i)$ (top), $Pr(P_i)$ (center) and $Res(P_i)$ (bottom), $i=0,2$, as $\beta_3$ varies, with $\beta_1>\beta_1^{th}$, which corresponds to the bifurcation diagram (ii) in Figure \ref{fig-bifb3} (top, right). As $\beta_3$ varies, we observe transitions between regimes III, II and I.}
\label{fig_NA_res_pre}
\end{figure}
\end{center}

We analyze the behavior of the above measures when parameters of system \eqref{sisNA} vary. Figure \ref{fig_NA_res_pre} shows the results when $\beta_3$ varies while other parameters are kept constant, with $\beta_1>\beta_1^{th}$, which corresponds to bifurcation diagram (ii) in Figure \ref{fig-bifb3} (center), where a transition I-II-III is observed. Results when other parameters vary are similar (not shown here).

To discuss these results, we first consider the point of view of cancer onset and analyze the resilience measures of $P_0$. In region I, $P_0$ is globally stable and all these measures are equal to the unity. When $\beta_3$ becomes lesser than $\beta_{3,\Delta}^{th}$ and enters region II, equilibrium $P_0$ is no longer globally stable, and its resilience measures undergo an abrupt jump and decay rapid in a small strip near $\beta_{3,\Delta}^{th}$. The most notorious jump occurs with $Pr(P_0)$. For values in the midpoint between the two thresholds separating region II from regions I and III, the values of $Pr(P_0)$ and $L(P_0)$ are very small. These features are due to the up-concave shape of the graphs of $L(P_0)$, $Pr(P_0)$ and $Res(P_0)$, and indicate that in the bistable regime the healthy state $P_0$ is threatened by small disturbances which may easily drive the system to the basin of attraction of $P_2$.

On the other hand, let us consider the point of view of treatment, and discuss the results concerning $P_2$. In regime III, this equilibrium is globally stable and $L(P_2)$, $Pr(P_2)$ and $Res(P_2)$ are equal to the unity. As $\beta_3$ becomes larger than $\beta_3^{th}$, these measures decay very slowly and are greater than the respective measures of $P_0$, until $\beta_3$ reaches the small strip near the next threshold, $\beta_{3,\Delta}^{th}$. Now, these features are due to the down-concave shape of the graphs, and implicate that $P_2$ is relatively protected against small perturbations.

By comparing these differences on the resilience measures of $P_0$ and $P_2$ we conclude that it is much more easy to drive the system outside the basin of attraction of the cure equilibrium $P_0$ when it loses its global stability, than driving the system out of the basin of attraction of the cancer equilibrium $P_2$ when it reaches the bistable regime, unless the parameters get very near the next bifurcation threshold at which $P_2$ loses its stability. In other words, our analysis reveals that, in the bistable regime, although these are different phenomena, it is more likely that mutations or exposure to carcinogenic factors drive cancer onset than chemotherapy, surgery or radiotherapy lead to tumor regression.

\subsection{The potential of resilience analysis to personalize cancer treatments}

We now illustrate how the resilience analysis can be used to obtain 
quantitative measures from models and use them as indicators to design personalized treatments.

\begin{center}
\begin{figure}[htb!]
\begin{minipage}[b]{\linewidth}
\begin{center}
\includegraphics[width=.327\linewidth]{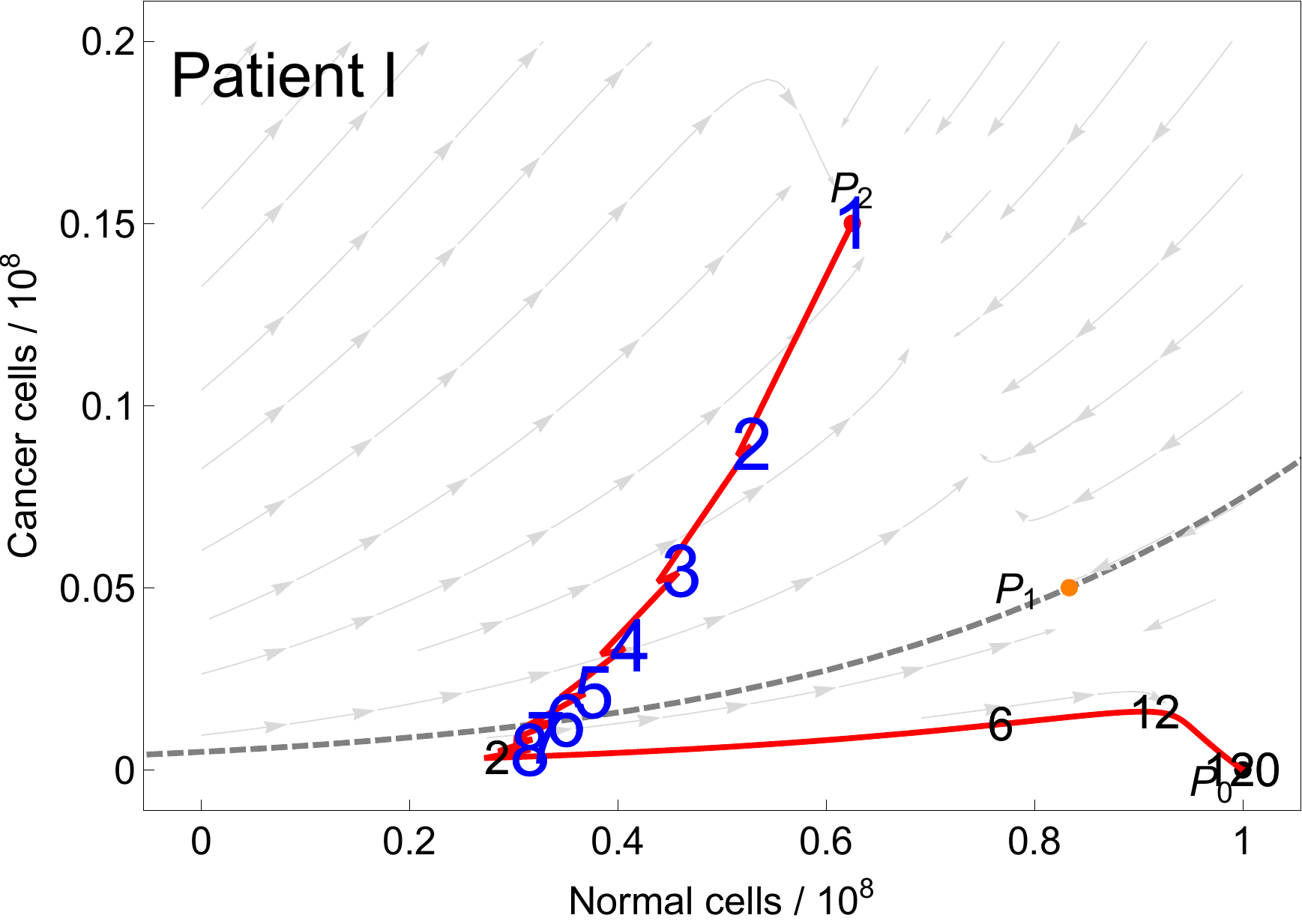}
\includegraphics[width=.327\linewidth]{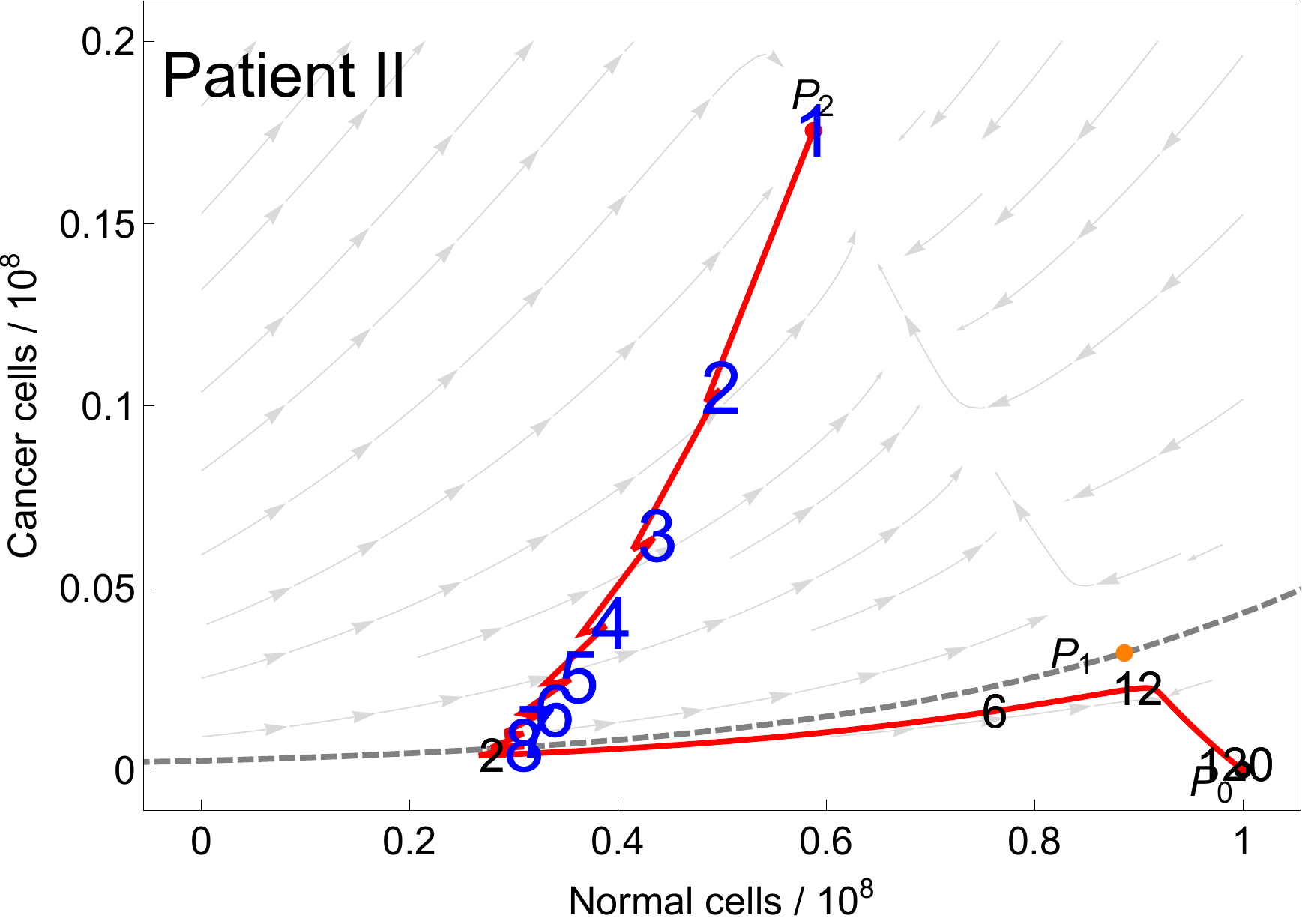}
\includegraphics[width=.327\linewidth]{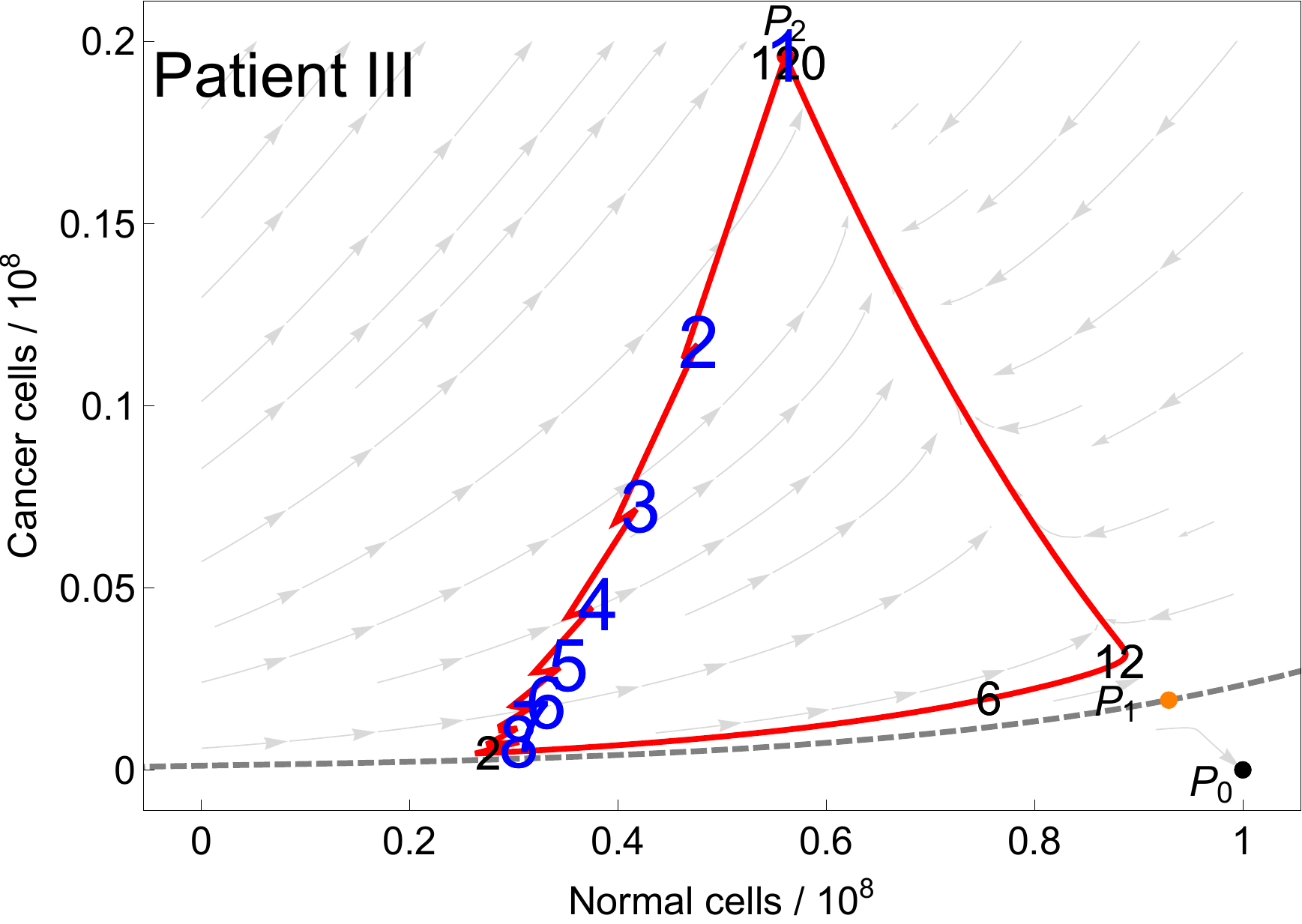}

\includegraphics[width=.327\linewidth]{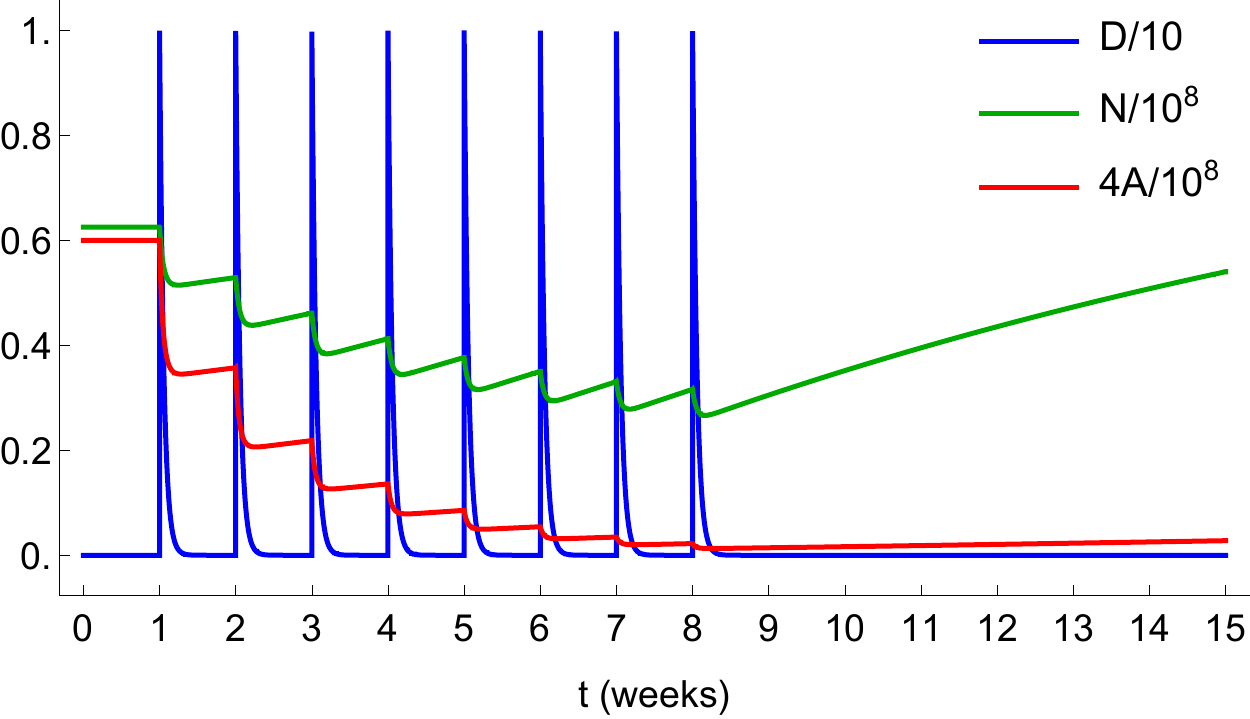}
\includegraphics[width=.327\linewidth]{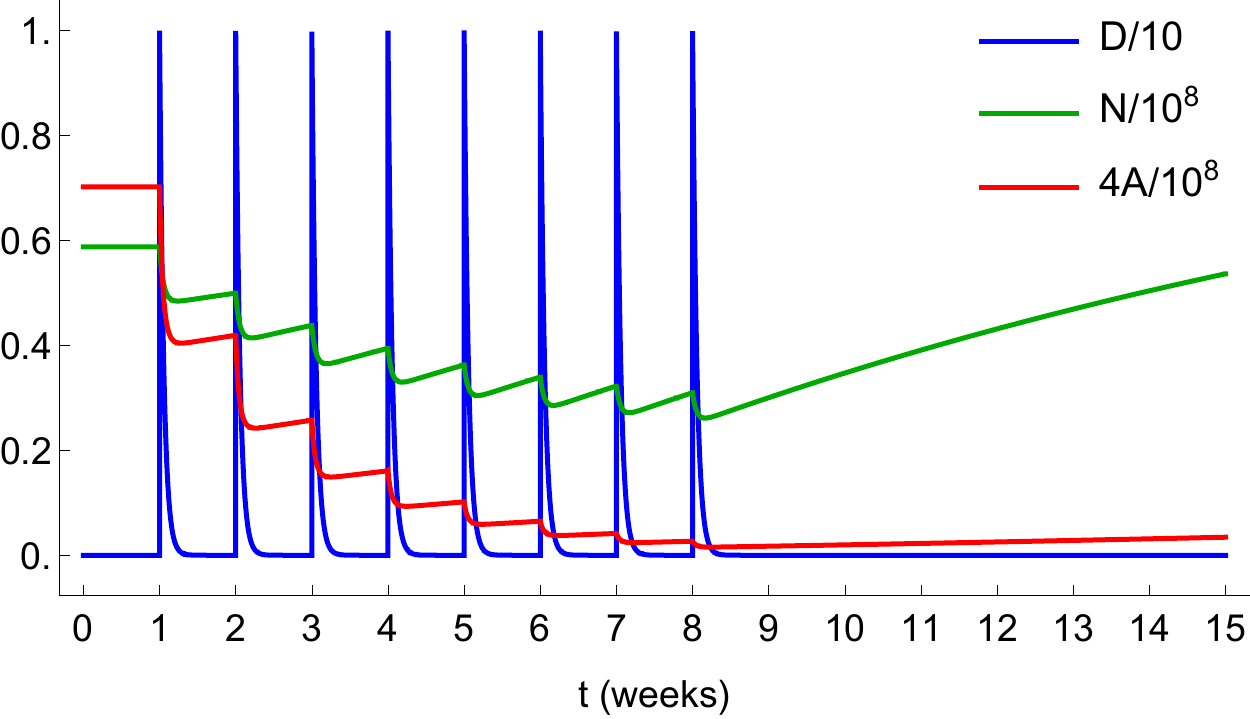}
\includegraphics[width=.327\linewidth]{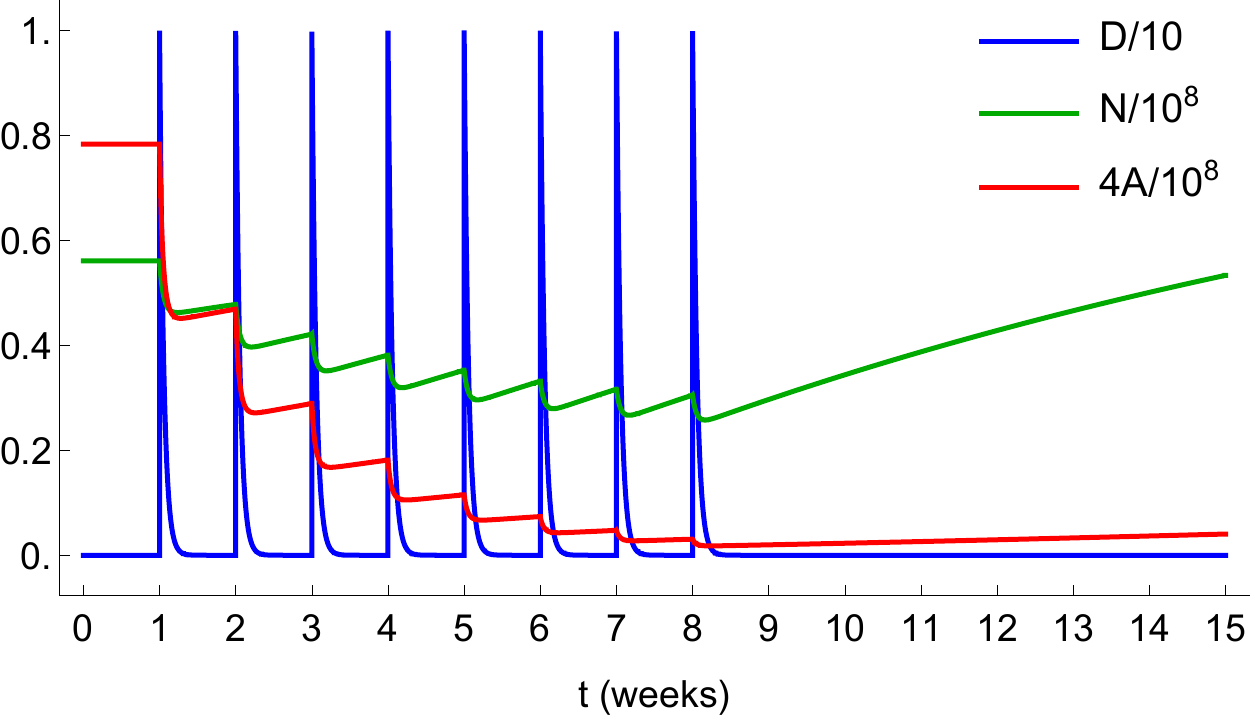}

\includegraphics[width=.327\linewidth]{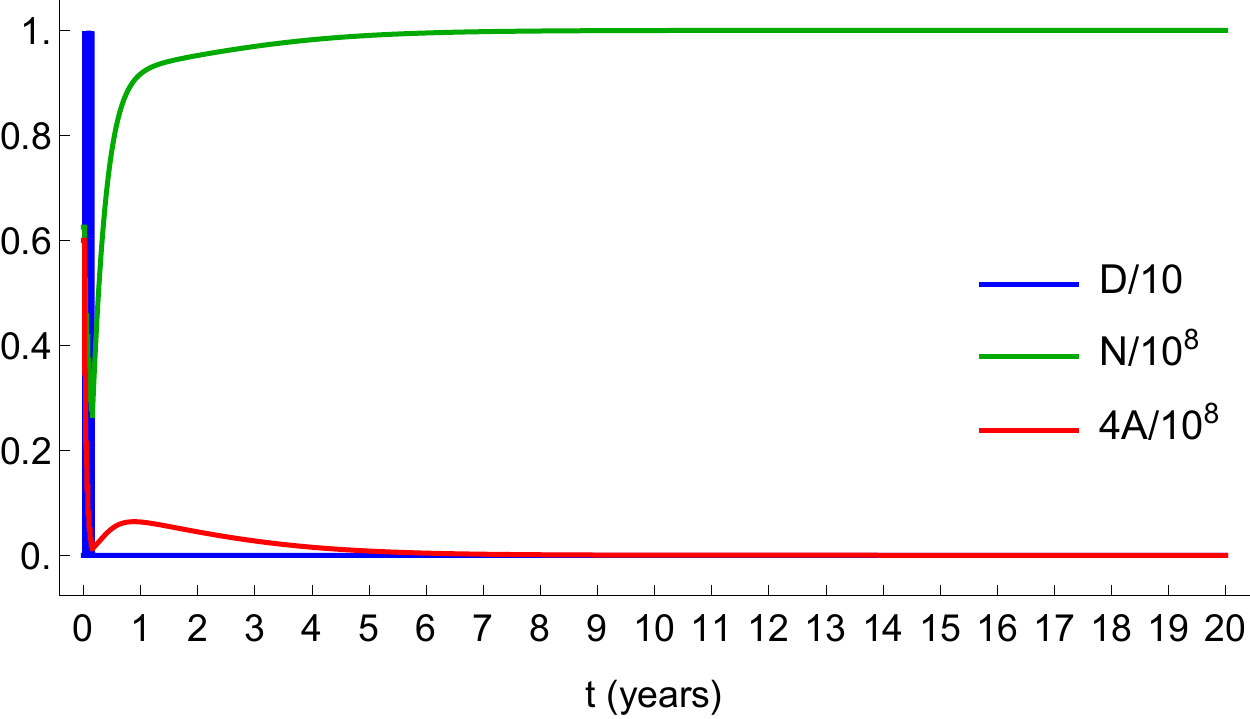}
\includegraphics[width=.327\linewidth]{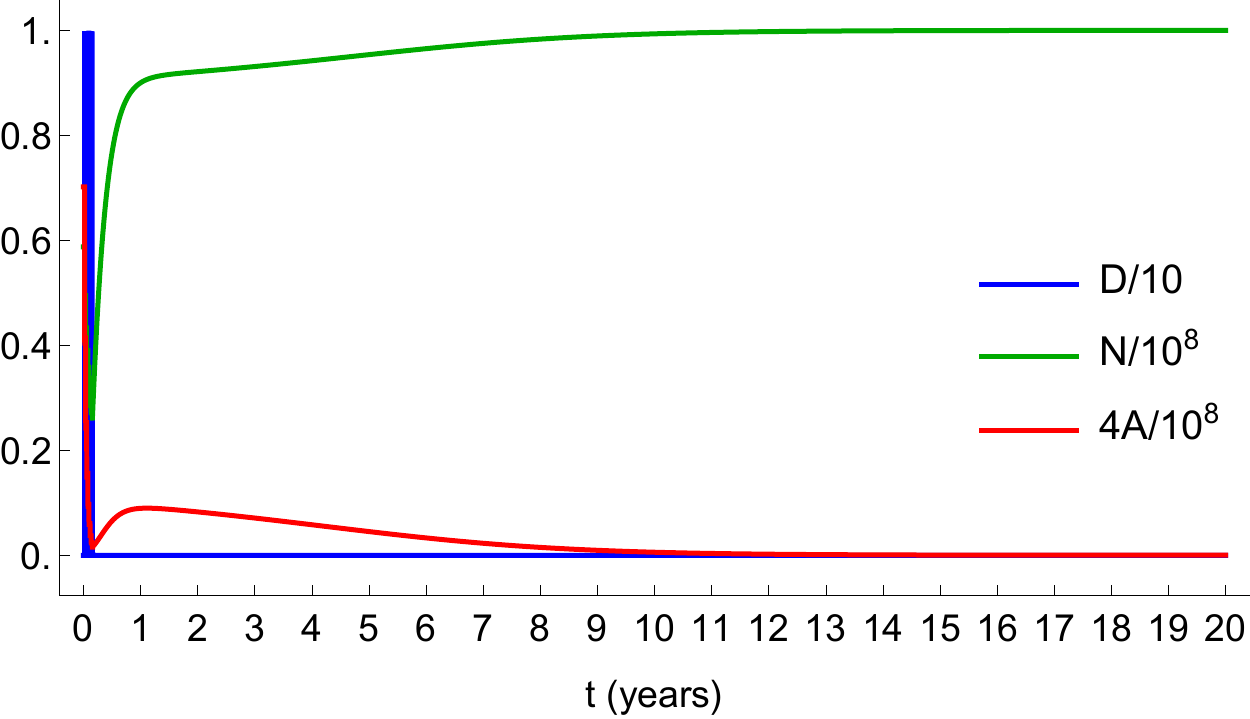}
\includegraphics[width=.327\linewidth]{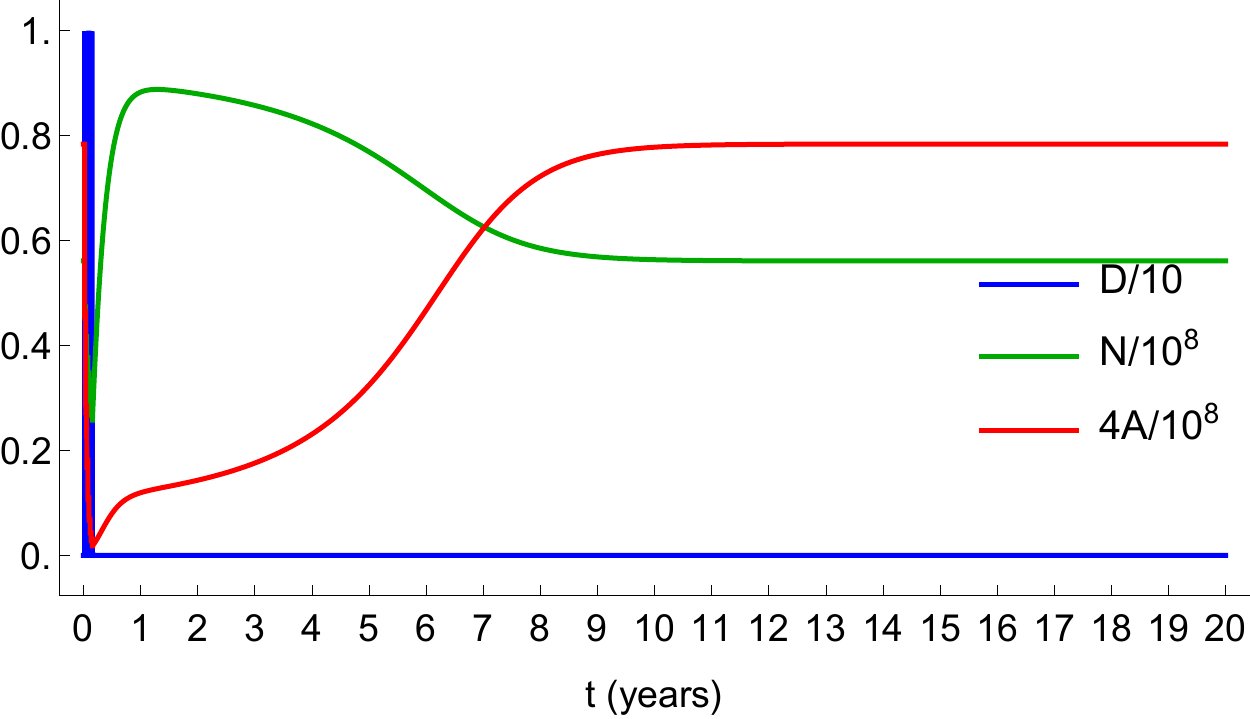}
\end{center}
\end{minipage}
\caption{Top row: projection of solutions of system \eqref{sisNAD_NA} in the $N\times A$ plane. The small black (large blue) numbers indicate the time in months (weeks) in which the solution was at each point. Center and bottom rows: short and long-term dynamics of solutions.}
\label{figura1}
\end{figure}
\end{center}

Figure \ref{figura1} illustrates how small differences in parameters lead to differences in the basins sizes and in the effectiveness of treatments. This Figure shows simulations of system \eqref{sisNAD_NA} for three almost identical individuals. The unique difference between them is the tumor apoptosis rate $\epsilon_A$. All them are treated with the same drug schedule, which consists in 8 weekly doses of $10$mg each. Patient I has the highest tumor apoptosis rate, $\epsilon_A=\epsilon_A^I=1/100$ day$^{-1}$. The treatment is completely effective since it ends in the `cure basin'. Further, there is a very small risk of tumor relapse, since small perturbations in this trajectory would not be able to drive it again to the tumor basin. Patient II has a smaller tumor apoptosis rate, $\epsilon_A=0.95 \epsilon_A^I$. The treatment is effective, but, once it ends very near the separatrix, there is a high risk of tumor relapse. Finally, patient III has the smallest tumor apoptosis rate $\epsilon_A=0.9\epsilon_A^I$. The treatment almost reaches the basin boundary but is not able to cross it. Thus, despite the number of tumor cells attains a low value at the end of the treatment, tumor relapse is observed after some time.

It is worth to note that the post-treatment situation of all three patients is almost the same. Indeed, according to the panels in the second and third row of Figure \ref{figura1}, the short term dynamics is quite similar for all the three patients during a transient period of almost one year. Only one or two years after the treatment the tumor of patient III starts to exhibit a substantial difference from the others. This is the same observed in many clinical cases. Once more, these results suggest that, from this theoretical point of view, those patients who presented tumor relapse could be completely cured if they were treated with a more intense drug schedule, which could be previously estimated based on some more complete information about them and their tumors.

\begin{center}
\begin{figure}[htb!]
\begin{minipage}[b]{\linewidth}
\begin{center}
\includegraphics[width=.325\linewidth]{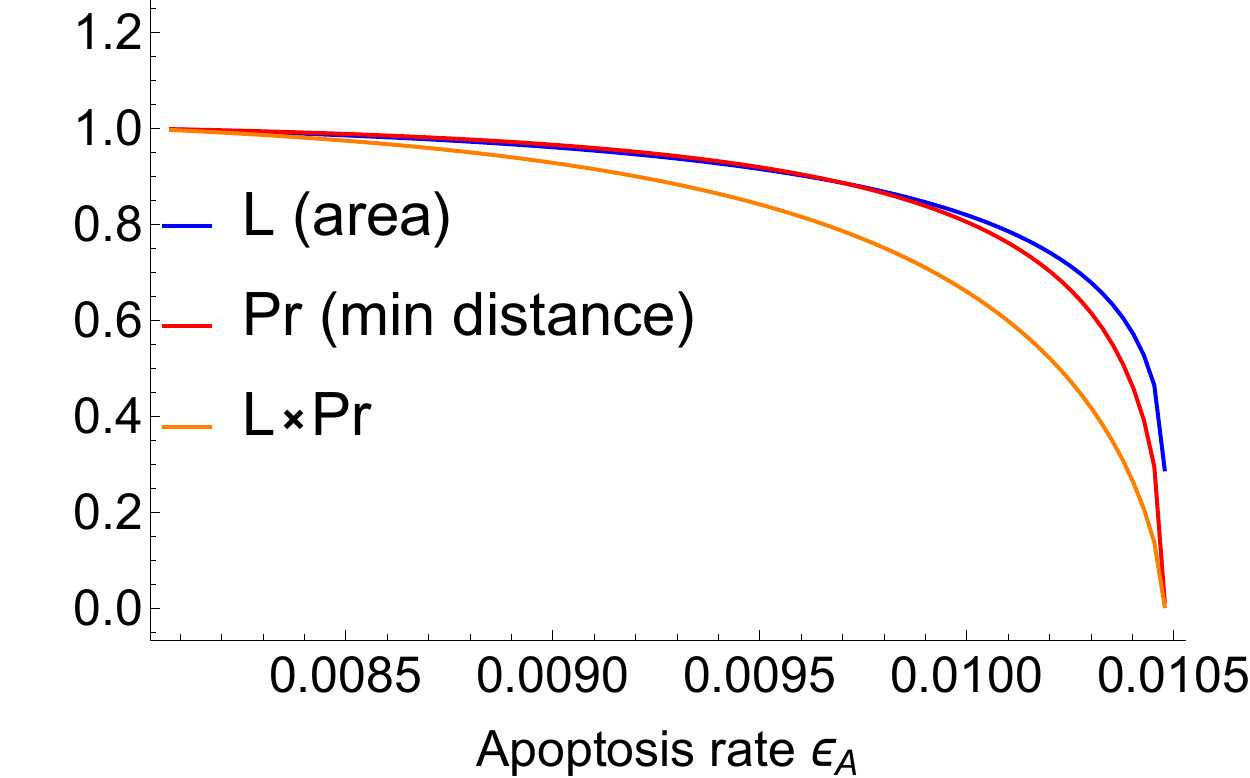}
\includegraphics[width=.325\linewidth]{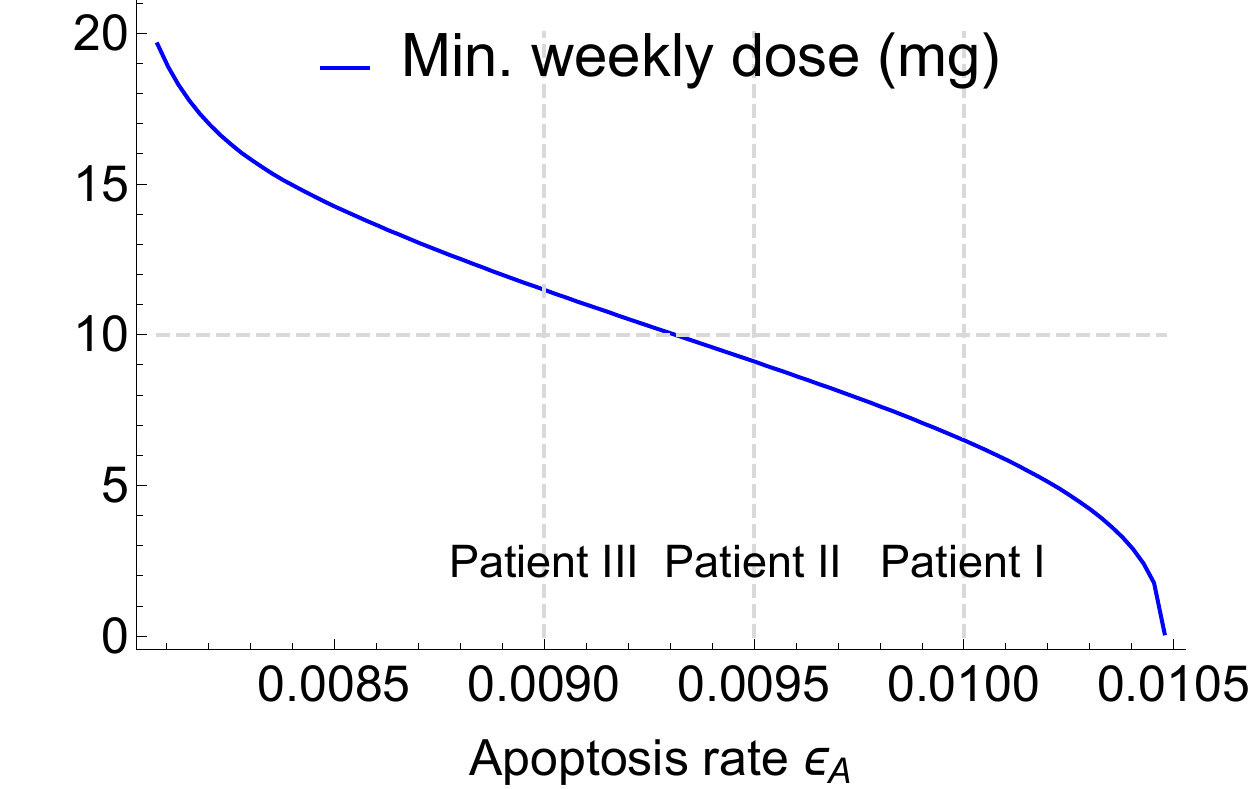}
\includegraphics[width=.325\linewidth]{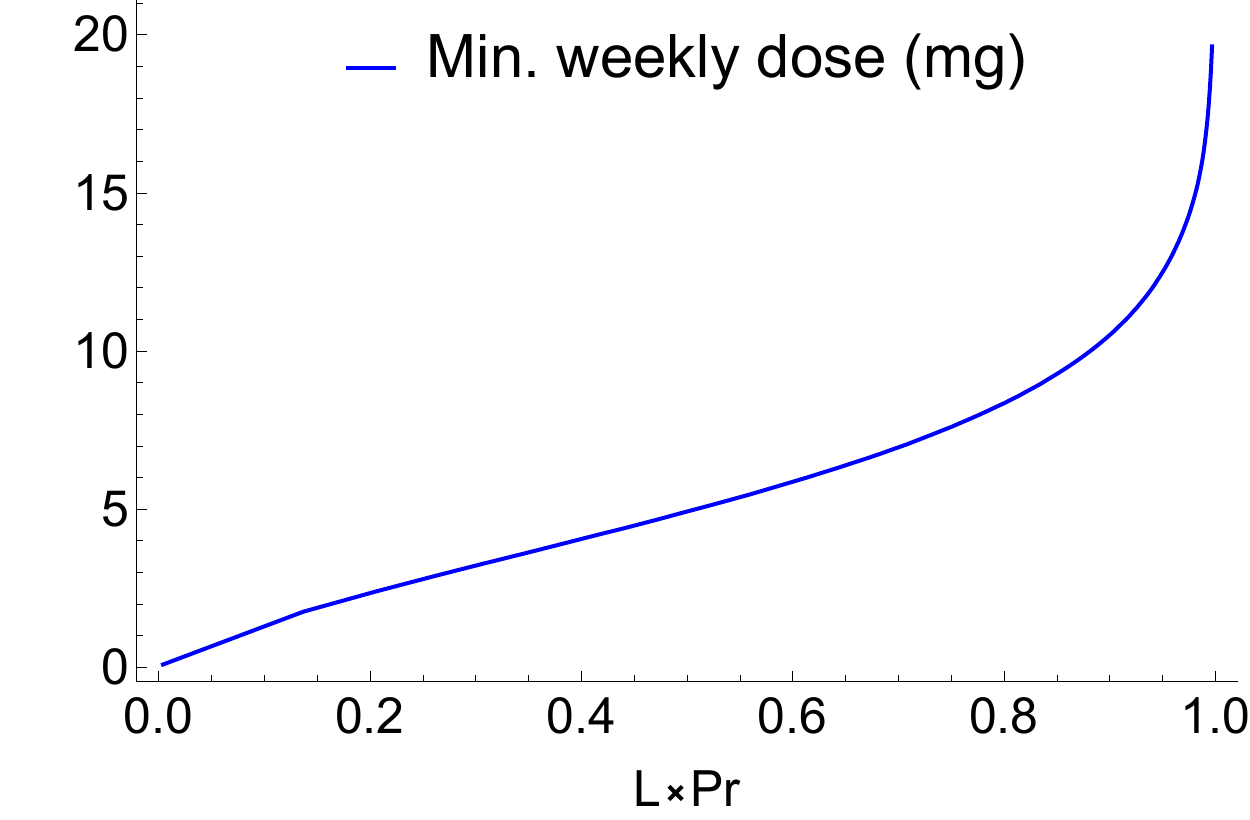}
\end{center}
\end{minipage}
\caption{Left: the behavior of resilience measures as $\epsilon_A $ varies. Center and right: behavior of the minimum weekly dose needed to enable an eight-week treatment to cross the boundary, as a function of $\epsilon_A$ (center) and the tumor resilience \textit{latitude} $\times$ \textit{longitude} (right).}
\label{figura2}
\end{figure}
\end{center}

The left panel of Figure \ref{figura2} shows how the \textit{latitude} and the \textit{precariousness} of the tumor basin, together with their product, behave when parameter $\epsilon_A$ varies. We see that the lower is the tumor apoptosis rate, the higher is the tumor resilience and the more difficult is to reach the cure basin. The center panel shows how the value of $\epsilon_A$ changes the minimum weekly dose needed to enable an eight-week treatment cross the boundary. This value varies in a large range, from $0$ mg to $20$ mg, $100\%$ with respect to the interval center ($10$mg), while $\epsilon_A$ varied in the interval $[0.8 \times 10^{-2},1.05 \times 10^{-2}]$, which is a small relative variation, about $\pm 13$\% with respect to the interval center, $\epsilon_A=0.925 \times 10^{-2}$. All graphs in this Figures \ref{figura1} and \ref{figura2} were obtained with the same parameters values from Table \ref{tablePar}, except $\epsilon_A$. We observe that a small decrease in $\epsilon_A$, of $10$\%, from $0.010$ to $0.009$, increases this minimum needed dose in $60$\%, from $7.5$ mg to $12$ mg. We also see that, in theory, if patient III receive a $12$ instead of a $10$ mg of drug dose each week, he would be cured. In case of toxicity constraints, an additional week of treatment would result in the same effect. Finally, the panel on the right shows the relationship between the `tumor resilience' (values of \textit{latitude} $\times$ \textit{longitude}) and the minimal needed dose.

Despite being obtained with a simple model for tumor growth which does not consider several important interactions in the tumor microenvironment, the results above show the potential of this kind of `resilience analysis` as a method to obtain indicators for design of tailored treatments design. The application of this approach to a validated model has the potential to elaborate a computational method to estimate the treatment needs for a particular patient or, at least, to stratify patients in a more refined fashion. The method would have as input data taken from the patient's tumor (its growth data obtained \textit{in vivo} or \textit{in vitro}). Then, a black-box function would calculate the total resilience of that tumor. This output would be used to prescribe a tailored treatment protocol which would be, in theory, effective to reach the cure of that patient.  The resilience approach was already applied to several other areas, but, as far as we know, this work is the first to do this for cancer.

\section{Conclusion}
\label{secNA:con}

In this paper, an ecological resilience framework to think of cancer as the alternance between two states was presented. This framework was based on the analysis of a simple ODE model for tumor growth considering the interaction with the host tissue. Despite the simplicity of the model, the approach adopted here gives interesting theoretical insights that shed some light on several relevant issues concerning cancer onset and treatment, and may help to improve the way we view cancer.

The model exhibited three regimes. These regimes were used to illustrate three different possibilities which may occur in clinical cases in general. The first regime corresponds to a healthy person where cancer onset is not possible since the cancer cure state is globally stable. The second regime corresponds to a person which can develop cancer if exposed to external carcinogenic factors, due to a partially corrupted repair system and/or a high aggressive phenotype of tumor cells. This regime presents bistability between the cancer and the cancer cure states. The third regime corresponds to a person in which cancer will arise due to intrinsic factors, i.e., the total corruption of repair systems. In this regime, the cancer state is globally stable.

Based on the general property that treatments are finite and, therefore, do not change the global dynamics in the phase space, we discussed the possibility of cure, which concerns stability and resilience questions above all. In the bistable regime the cure is possible if the treatment is able to drive the system to the basin of attraction of the cure equilibrium. Tumor recurrence in this case is associated with treatments which are unable to cross the separatrix between the basins, or which do not end at a safe distance from the separatrix. In the third regime, the complete cure is not possible at all, since the repair system is intrinsically weak, but tumor recurrence may be delayed if the treatment is prolonged, because the system takes a long time to pass around the cure equilibrium, which is a saddle point. However, toxicity was not assessed in this model. 

Besides perturbations on state variables, a view in the switching between these three regimes due to parameters changes in a slow time scale was discussed and the roles of the most important parameters in these transitions were assessed. Results indicated that only aggressive tumors may arise  if intrinsic repair systems are not totally corrupted. Further, these aggressive tumors depend on exposure to external carcinogenic factors for arising.

Finally, we reviewed the use of three different measures to assess the resilience of a stable equilibrium. We proposed a simple and efficient methods to calculate these measures. After applying this analysis to the model we concluded that in the bistable regime the cancer equilibrium has much more resilience than the cure equilibrium, with respect to state variable perturbations as parameters changes. We also illustrated how resilience analysis can be used to estimate the treatment needs for a particular patient, and showed its potential to design personalized cancer treatments.

This paper contributes to the current understanding on cancer by raising some issues in an ecological view, and also demonstrates how a `resilience analysis' may be applied to population dynamics models in order to improve the understanding of nonlinear phenomena. Further, the application of the `resilience analysis' presented here to detailed and accurate models for specific tumor types has the potential to generate measures which can be accounted for in the design of treatment plans for cancer patients and also in the development of adaptive treatments \cite{benzekry2015metronomic}. The size and shape of the basin of attraction of the cancer equilibria in those models may be used as indicators in the design and planning of personalized treatments.

\section{Acknowledgements}

The authors would like to thank to two anonymous referees, whose suggestions contributed to improve the quality of the paper.

A. C. F. was partially supported by FAPEMIG (Fundação de Amparo à Pesquisa do Estado de Minas Gerais), process PEE-00887-16.

\printbibliography

\end{document}